\documentclass[reprint,superscriptaddress,amsmath,amssymb,
 aps,showkeys]{revtex4-2}
\usepackage{graphicx}
\usepackage{dcolumn}
\usepackage{bm}
\usepackage{multirow}
\usepackage{booktabs}
\usepackage{float}
\usepackage{placeins}
\usepackage{xcolor}
\usepackage[colorlinks=true,linkcolor=blue,citecolor=blue,urlcolor=blue]{hyperref}
\usepackage{miller}

\begin{document}

\preprint{APS/123-QED}

\title{Predicting Grain Boundary Segregation in Magnesium Alloys: \\An Atomistically Informed Machine Learning Approach}

\author{Zhuocheng Xie}
\email[]{xie@imm.rwth-aachen.de}
\affiliation{Institute of Physical Metallurgy and Materials Physics, RWTH Aachen University, 52056 Aachen, Germany}
    
\author{Achraf Atila}
\email[]{achraf.atila@bam.de}
\affiliation{Department of Materials Science and Engineering, Saarland University, 66123 Saarbrücken, Germany}
\affiliation{Federal Institute of Materials Research and Testing (BAM), Unter den Eichen 87, Berlin 12205, Germany}

\author{Julien Guénolé}
\affiliation{CNRS, Université de Lorraine, Arts et Métiers, LEM3, 57070 Metz, France}

\author{Sandra Korte-Kerzel}
\affiliation{Institute of Physical Metallurgy and Materials Physics, RWTH Aachen University, 52056 Aachen, Germany}

\author{Talal Al-Samman}
\affiliation{Institute of Physical Metallurgy and Materials Physics, RWTH Aachen University, 52056 Aachen, Germany}

\author{Ulrich Kerzel}
\affiliation{Institute of Physical Metallurgy and Materials Physics, RWTH Aachen University, 52056 Aachen, Germany}

\begin{abstract}
Grain boundary (GB) segregation substantially influences the mechanical properties and performance of magnesium (Mg). Atomic-scale modeling, typically using ab-initio or semi-empirical approaches, has mainly focused on GB segregation at highly symmetric GBs in Mg alloys, often failing to capture the diversity of local atomic environments and segregation energies, resulting in inaccurate structure-property predictions. This study employs atomistic simulations and machine learning models to systematically investigate the segregation behavior of common solute elements in polycrystalline Mg at both 0 K and finite temperatures. The machine learning models accurately predict segregation thermodynamics by incorporating energetic and structural descriptors. We found that segregation energy and vibrational free energy follow skew-normal distributions, with hydrostatic stress, an indicator of excess free volume, emerging as an important factor influencing segregation tendency. The local atomic environment’s flexibility, quantified by flexibility volume, is also crucial in predicting GB segregation. Comparing the grain boundary solute concentrations calculated via the Langmuir–McLean isotherm with experimental data, we identified a pronounced segregation tendency for Nd, highlighting its potential for GB engineering in Mg alloys. This work demonstrates the powerful synergy of atomistic simulations and machine learning, paving the way for designing advanced lightweight Mg alloys with tailored properties.
\end{abstract}


\maketitle

\section{Introduction}

Magnesium (Mg) alloys are lightweight structural materials with promising applications in sustainable, low-carbon industries, despite challenges with limited formability and ductility due to strong crystallographic texture \cite{mordike2001magnesium,pollock2010weight,yoo1981slip,kocks2000texture,song2020latest,prasad2022role,bai2023applications}. Addressing these issues involves alloying to balance critical resolved shear stresses \cite{sandlobes2011role,wang2022optimizing,ovri2023mechanistic} and leveraging solute grain boundary (GB) segregation to influence texture \cite{basu2014triggering,trang2018designing,chen2022mechanisms}. Solute segregation at GBs has been widely reported to influence GB energy and mobility, subsequently affecting the texture development of Mg alloys~\cite{zeng2016texture,barrett2017effect}. Understanding the atomistic origins of GB segregation phenomena is crucial for designing Mg alloys with specifically tailored properties.

Nie et al.~\cite{nie2013periodic} reported periodic substitutional segregation of Gd and Zn solutes at twin boundaries in Mg alloys using high-resolution scanning transmission electron microscopy and attributed this periodic solute decoration to strain energy minimization using the density functional theory (DFT) calculations. The pinning effect of this ordered solute segregation on twin boundaries leads to the annealing strengthening of these Mg alloys. Similar atomic-resolution electron microscopic approaches have been carried out on other highly symmetric tilt GBs in Mg alloys, where characteristic solute decoration patterns were similarly observed~\cite{zhou2015effect,he2021unusual,xie2021nonsymmetrical}. Extensive experimental efforts have been made to quantify solute concentrations at GBs in Mg alloys. This was driven by the fact that certain alloying elements (rare earth, Y, and Ca) exhibit a pronounced tendency to segregate at these boundaries, even at low concentrations within the Mg matrix, involving intricate atomic-scale interactions~\cite{pei2021grain,fu2022achieving,li2022elucidation,li2022elucidating,pei2022synergistic,yi2023interplay,zhang2023anisotropic,pan2020mechanistic,basu2022segregation,pei2022effect,hadorn2012role,robson2016grain,langelier2017effects,qian2022influence,zhang2022significantly,pei2023atomistic,mouhib2024exploring}. Pei et al.~\cite{pei2023atomistic} measured varying Nd solute concentrations between 2 and 5 at.\%, across six different GBs in Mg alloys using atom probe tomography (APT). The observed inhomogeneous segregation behavior within the GB plane, which was rationalized through correlated atomistic simulations on general GBs with the same misorientations and GB planes, suggests that it stems from local atomic arrangements within the GBs rather than macroscopic characteristics. Using APT, Mouhib et al.~\cite{mouhib2024exploring} characterized the concentrations of Ca and Gd solutes at GBs in Mg alloys, both with and without adding Zn. The selective formation of specific texture components in the ternary Mg-Zn-X alloys (X= Ca or Gd) was correlated to the synergistic effects of Zn addition on solute clustering and binding via DFT.

Atomic-scale modeling techniques, including DFT and atomistic simulations, have been applied to deepen our fundamental understanding of GB segregation in Mg alloys. Huber et al.~\cite{huber2014atomistic} calculated the per-site segregation energy $\Delta E_{\text{seg}}$ of 11 alloying elements to $\Sigma$7 GBs in Mg using DFT. They developed a linear elastic model based on the GB site volume and matrix's bulk modulus that accurately captures the segregation behavior of solute elements in highly symmetric GBs due to strain energy minimization. Wang et al.~\cite{wang2024defects} systematically assessed Mg-X semi-empirical potentials for $\Delta E_{\text{seg}}$ of alloying elements (X= Al, Ca, Li, Sn, Y, Nd, Pb, and~Zn) at compression and tensile twin boundaries as well as $\Sigma$7 GBs in Mg. The calculated $\Delta E_{\text{seg}}$ from these interatomic potentials are in qualitative agreement with DFT calculations~\cite{huber2014atomistic,pei2019first} and demonstrate a strong correlation with the site volume. 
Wagih et al.~\cite{wagih2020learning} calculated $\Delta E_{\text{seg}}$ spectra of multiple solute elements at GBs in a polycrystalline Mg sample (with dimensions of 20 $\times$ 20 $\times$ 20 nm$^{3}$ consisting of 16 grains) using different semi-empirical potentials. They performed linear regression machine learning (ML) on the structural features generated using Smooth Overlap of Atomic Positions (SOAP) to correlate with the skew-normal $\Delta E_{\text{seg}}$ distribution.
Messina et al.~\cite{messina2021machine} employed various regression ML models to predict $\Delta E_{\text{seg}}$ of Al solute at Mg \hkl<0001> symmetric tilt GBs, calculated using atomistic simulations to the energetic and structural descriptors (ESD), where the XGBoostRegressor model demonstrated the best performance with a $R^{2}$ score of 0.972. Recently, Menon et al.~\cite{menon2024atomistic} calculated $\Delta E_{\text{seg}}$ of Y solute at symmetric tilt GBs in Mg using atomistic simulations and employed the thermodynamic integration method to estimate the per-site segregation free energy $\Delta F_{\text{seg}}$ at finite temperatures. Based on the Langmuir–McLean model~\cite{mclean1957grain} and the extension by White and Coghlan~\cite{white1977spectrum} utilizing the calculated $\Delta F_{\text{seg}}$ spectra, the predicted GB concentrations of Y solute showed a good correlation with experimentally measured concentrations at typical thermomechanical processing temperatures in Mg alloys. Further, they trained a stacking cross-validation regression ML model on hundreds of data points using physics-informed descriptors to predict $\Delta F_{\text{seg}}$.

Atomic-scale simulations via ab-initio or semi-empirical approaches have mostly focused on GB segregation at highly symmetric GBs in Mg alloys, which may not sufficiently capture the broad spectra of local atomic environments and segregation thermodynamic quantities at GBs, leading to inaccurate predictions in structure-property relationships~\cite{wagih2023can}. Although Wagih's work~\cite{wagih2020learning} covers a wide range of solute elements in polycrystalline Mg, the differing interatomic potentials with different Mg-Mg interactions make comparisons difficult, particularly since the embedded-atom method (EAM) potentials~\cite{zhou2004misfit} were not developed for modeling hexagonal close-packed Mg. Additionally, most solute elements studied are rarely used in newly developed Mg-based solid solutions for practical, lightweight applications. Therefore, no experimental data are available for comparison with the theoretical and computational approaches. To date, there has been no comprehensive investigation of the GB segregation spectrum across various common solute elements in polycrystalline Mg at both absolute zero and finite temperatures.
In this study, we systematically investigated GB segregation of six solute elements---Nd, Ca, Y, Li, Al, and Zn---commonly used in wrought Mg alloy sheets and extrusions. Our investigations employed atomistic simulations within a magnesium polycrystal framework comprising a broad variety of site types of the full GB space. In addition to determining the segregation energy spectra at 0~K, we also accounted for the vibrational entropy contribution at finite temperatures using the harmonic approximation (HA) approach. The predicted GB solute concentrations were directly compared with experimental data from the literature, covering relevant bulk concentrations and temperature regimes. Furthermore, ML approaches were employed in combination with atomistic modeling to predict both the segregation energy and the segregation free energy, with the goal of providing rapid predictions using the essential structural and energetic parameters within the pure Mg polycrystal at a reduced computational cost.

\section{Methods} \label{methods}
\subsection{Atomistic simulations}
Atomistic simulations in this study were performed using the molecular dynamics (MD) software package LAMMPS~\cite{thompson2022lammps}. 
Interatomic interactions were modeled by the modified embedded atom method (MEAM) potentials developed by Lee et al. for Mg--X (X= Nd~\cite{kim2017modified}, Ca~\cite{kim2015modified}, Y~\cite{kim2015modified}, Li~\cite{kim2012atomistic}, Al~\cite{kim2009atomistic} and Zn~\cite{jang2018modified}) systems, where the Mg--Mg interaction~\cite{kim2009atomistic} is identical. These MEAM potentials have shown good performance in predicting substitutional segregation energies at highly symmetric GBs in Mg~\cite{wang2024defects}.

The polycrystalline Mg structure with a random texture was constructed via Voronoi tessellation using Atomsk~\cite{hirel2015atomsk}. The periodic simulation box, measuring 15 nm in dimensions, contains 12 grains and approximately 145,000 atoms, with 26 \% of these sites characterized as GB sites according to the adaptive common neighbor analysis (CNA)~\cite{stukowski2012structure}. Following initial energy minimization, the sample was heated to 500 K over 50 ps and annealed at 500 K for 20 ps before being quenched back to 0 K over 50 ps using the isothermal-isobaric ensemble with the Nosé-Hoover thermostat and barostat~\cite{hoover1985canonical}. Subsequently, the FIRE algorithm~\cite{bitzek2006structural,guenole2020assessment} was applied to relax the structure with a force tolerance of $10^{-8}$ eV/$\text{\AA}$.

The flexibility volume ($V_{\text{flex}}$)~\cite{ding2016universal} of a GB site in the polycrystal was calculated following Equation~\eqref{equ0}:
\begin{equation}
\label{equ0}
V_{\text{flex}} = \left<MSD_{\text{vib},i}\right> \sqrt[3]{\Omega_{i}},
\end{equation}
where $\Omega_{i}$ is the atomic volume of the $i$th atom according to the Voronoi tessellation. The vibrational mean-squared displacement ($MSD_{\text{vib},i}$) of the $i$th atom was calculated using $\left[ x_{i}(t_0)-x_{i}(t_0+\Delta t) \right]^2$, where $x_{i}(t_0)$ is the equilibrium position of the $i$th atom and  $x_{i}(t_0 + \Delta t)$ is its position after 5000 MD steps (timestep is 1 fs). The system was initialized by assigning velocities corresponding to an initial temperature of 600 K and then simulated under the microcanonical (NVE) ensemble. Due to the equipartition of the potential energy and kinetic energy, the energy fluctuates around half of the initially assigned temperature value and effectively maintains the temperature near 300 K during the simulation. The overall MSD of the polycrystalline system is flat with time and contains the vibrational but not the diffusional contribution, see Figure S1 (Supplementary Material).
The averaged vibrational mean-squared displacement $\left<MSD_{\text{vib},i}\right>$ was calculated by averaging over 100 independent MD runs with different initial velocity distributions.

\subsection{Segregation energy and vibrational entropy contribution}

The segregation energy $\Delta E_{\text{seg}}$ at 0 K for a substitutional solute atom X at a GB site was calculated using Equation~\eqref{equ1}:
\begin{equation}
\label{equ1}
\Delta E_{\text{seg}} = (E_{\text{bulk}} - E_{\text{bulk,X}}) - (E_{\text{GB}} - E_{\text{GB,X}}),
\end{equation}
where $E_{\text{bulk}}$ represents the energy of a bulk system, $E_{\text{bulk,X}}$ denotes the energy of the bulk system with a host atom replaced by an X solute, $E_{\text{GB}}$ is the energy of a system containing a GB, and $E_{\text{GB,X}}$ stands for the energy of the system with a solute atom X occupying the GB site. The structures were relaxed using the FIRE algorithm with a force tolerance of $10^{-8}$ eV/$\text{\AA}$. A negative segregation energy indicates favorable solute segregation.

The segregation free energy $\Delta F_{\text{seg}}$ of a substitutional solute in a solid at finite temperatures $T$ was computed by considering the contribution of entropy change (excess vibrational entropy) $\Delta S_{\text{seg}}$, as given by Equation~\eqref{equ2}:
\begin{equation}
\label{equ2}
\Delta F_{\text{seg}} = \Delta E_{\text{seg}} - T\Delta S_{\text{seg}}.
\end{equation}
In this study, among all excess entropy contributions, the vibrational entropy is the only significant contribution to $\Delta F_{\text{seg}}$. The segregation free energy can thus be written:
\begin{equation}
\label{equ3}
\Delta F_{\text{seg}} = \Delta E_{\text{seg}} + \Delta F_{\text{seg}}^{\text{vib}},
\end{equation}
with the segregation vibrational free energy $\Delta F_{\text{seg}}^{\text{vib}}$ calculated using Equation~\eqref{equ4}:
\begin{equation}
\label{equ4}
\Delta F_{\text{seg}}^{\text{vib}} = (F_{\text{bulk}}^{\text{vib}} - F_{\text{bulk,X}}^{\text{vib}}) - (F_{\text{GB}}^{\text{vib}} - F_{\text{GB,X}}^{\text{vib}}),
\end{equation}
where $F_{\text{bulk}}^{\text{vib}}$ represents the vibrational free energy of a bulk system, $F_{\text{bulk,X}}^{\text{vib}}$ denotes the vibrational free energy of the bulk system with a host atom replaced by an X solute, $F_{\text{GB}}^{\text{vib}}$ is the vibrational free energy of a system containing a GB, and $F_{\text{GB,X}}^{\text{vib}}$ stands for the vibrational free energy of the system with a solute atom X occupying the GB site.

To estimate the vibrational entropy contribution, the harmonic approximation (HA) was employed~\cite{creuze2000intergranular, tuchinda2023vibrational}, where full HA was performed for a spherical region centered at the GB site of interest with a chosen cutoff radius and boundary conditions as illustrated in Figure S2. A convergence test on the cutoff radius was conducted, confirming the selection of a cutoff radius of 14 $\text{\AA}$ (see Figure S2). The vibrational frequency $\nu$ was determined by the equation of motion:
\begin{equation}
\label{equ5}
M_{i} \ddot{u}_{i\alpha} = -\frac{\partial U}{\partial u_{i\alpha}} = -\sum_{j,\beta} \Phi_{i\alpha j\beta}^{0} u_{j\beta},
\end{equation}
where $M_{i}$ is the atomic mass of atom $i$, $u_{i\alpha}$ is the displacement of the $i$ atom in the direction $\alpha$, $U$ is the potential energy of the system, and $\Phi_{i\alpha j\beta}^{0}$ is the force constant of atom $i$ in the direction $\alpha$ when atom $j$ is displaced in the $\beta$ direction. Equation~\eqref{equ5} can be rewritten by looking for solutions of the form $u_{i\alpha}(t) = 1 / \sqrt{M_i} \tilde{u}_{i\alpha} \exp(2 \pi i \nu t)$: 
\begin{equation}
\label{equ6}
\nu^2 \tilde{u}_{i\alpha} = \sum_{j,\beta} D^{0}_{i\alpha j\beta} \tilde{u}_{j\beta},
\end{equation}
where $D^{0}_{i\alpha j\beta} = \Phi^{0}_{i\alpha j\beta} / 4\pi^2 \sqrt{M_i M_j}$ is a (3$N$ $\times$ 3$N$) dynamical matrix of the full HA region for the GB site with the total number of atoms $N$. $\tilde{u}$ is the mass-scaled displacement. The vibrational frequencies obtained by diagonalizing the dynamical matrix were then utilized to calculate the vibrational free energy $F^{\text{vib}}$ via Equation~\eqref{equ7}.
\begin{equation}
\label{equ7}
F^{\text{vib}} = k_\text{B} T \sum_{i=1}^{3N} \ln \left[ 2 \sinh \left( \frac{h \nu_i}{2 k_\text{B} T} \right) \right],
\end{equation}
where $k_\text{B}$ is the Boltzmann constant and $h$ is the Planck constant. The vibrational entropy $S^{\text{vib}}$ can be obtained: 
{\small
\begin{equation}
\label{equ8}
S^{\text{vib}} = -k_\text{B} \sum_{i=1}^{3N} \left[ \ln \left( 2\sinh \left( \frac{h\nu_i}{2k_\text{B}T} \right) \right) - \frac{h\nu_i}{2k_\text{B}T} \coth \left( \frac{h\nu_i}{2k_\text{B}T} \right) \right],
\end{equation}}\\
and the vibrational heat capacities at constant volume $C_{\text{v}}$ and constant pressure $C_{\text{p}}$ can be derived from $S^{\text{vib}}$:
\begin{equation}
\label{equ9}
C_{\text{v}} = T\left(\frac{\partial S^{\text{vib}}}{\partial T}\right)_{\text{v}},
\end{equation}
\begin{equation}
\label{equ10}
C_{\text{p}} = C_{\text{v}} + (2\alpha_a + \alpha_c)^2 BVT,
\end{equation}
where $\alpha_a$ and $\alpha_c$ are thermal expansion coefficients of $a$ and $c$ lattice constants, respectively. $B$ is bulk modulus and $V$ is molar volume.

\subsection{Machine learning prediction}
In this study, we applied ML to predict per-site segregation energy $\Delta E_{\text{seg}}$, segregation vibrational free energy $\Delta F_{\text{seg}}^{\text{vib}}$, and segregation free energy $\Delta F_{\text{seg}}$ at 600~K in binary Mg alloys using datasets representing six alloying elements: Nd, Ca, Y, Li, Al and Zn. Each dataset contained 38,114 GB sites and 16 energetic and structural descriptors (ESD), including potential energy, six stress tensors ($\sigma_{ij}$), hydrostatic stress ($\sigma_{h}$), coordination number, centrosymmetric parameter (CSP), structural features obtained from Voronoi tessellation (atomic volume, cavity radius, and surface area of the polyhedra), vibrational free energy $F^{\text{vib}}$ at 600 K, vibrational MSD ($MSD_{\text{vib}}$), and flexibility volume ($V_{\text{flex}}$). We preprocessed the data by selecting relevant features, excluding the target variables, and performing correlation analysis using heatmaps. In addition to ESD, we generated the fingerprint of the local atomic environment at GB sites using SOAP~\cite{bartok2010gaussian,bartok2013representing}. The SOAP approach models the Gaussian particle density functions for all neighboring atoms in the local atomic environment using a combination of radial basis functions and spherical harmonics. The parameters controlling the SOAP descriptor vector include the cutoff radius ($r_\text{max}$ = 6 \AA), the maximum number of radial basis functions ($n_\text{max}$ = 12), the degree of spherical harmonics ($l_\text{max}$ = 12), and the Gaussian function width ($\sigma_\text{at}$ = 1 \AA). In total, the SOAP vector for each GB site comprises 1015 features. The SOAP descriptors were combined with the ESD features to enhance prediction accuracy. For the prediction of $\Delta F_{\text{seg}}^{\text{vib}}$, $\Delta E_{\text{seg}}$ was included as an input variable along with other features. 

The CatBoostRegressor model~\cite{prokhorenkova2018catboost}, known for its ability to handle categorical data efficiently and its robust performance with minimal hyperparameter tuning, was chosen for this task. Each dataset was split into training, testing, and validation sets (70-20-10 ratio). The split in the learning algorithm allocated 70\% of the data to training, 20\% to testing, and 10\% to validation.
The hyperparameter space was explored using GridSearchCV with a predefined parameter grid, which included various combinations of iterations, learning rate, tree depth, L2 regularization, and bagging temperature, to identify the optimal settings that maximize the $R^{2}$ score. The model was trained with the optimized hyperparameters using 1000 iterations, a learning rate of 0.03, a tree depth of 8, an L2 regularization term of 1, and a bagging temperature of 0.5 to ensure convergence, see Figure S3(d). 5-fold cross-validation was performed on the training set to evaluate the model’s robustness and stability.
Model performance was evaluated on both the testing and validation sets using mean absolute error (MAE) and $R^{2}$ scores. We visualized the results through scatter plots of true versus predicted values with Gaussian Kernel Density Estimation (KDE) to highlight prediction densities, histograms comparing the distributions of true and predicted per-site energies, and feature importance plots.

To assess the performance of the proposed ML model, we evaluated GradientBoostingRegressor~\cite{friedman2001greedy}, XGBoostRegressor~\cite{chen2016xgboost}, and ElasticNetCV~\cite{zou2005regularization} models implemented in scikit-learn~\cite{pedregosa2011scikit} in addition to CatBoostRegressor on the same datasets consisting of the ESD features using the default hyperparameters. The evaluation metrics demonstrated that all models showed similar behavior, but CatBoostRegressor slightly outperformed the others, achieving the lowest MAE values and the highest $R^{2}$ scores across all datasets in predicting $\Delta E_{\text{seg}}$ in the investigated Mg alloys, making it the most reliable model for this application. The benchmarking results are compiled in Table S1 and Figure S3, providing a comprehensive comparison of model performances.

\section{Results}

\subsection{Thermal properties of pure Mg}
\begin{figure*}[hbt!]
\centering
\includegraphics[width=\textwidth]{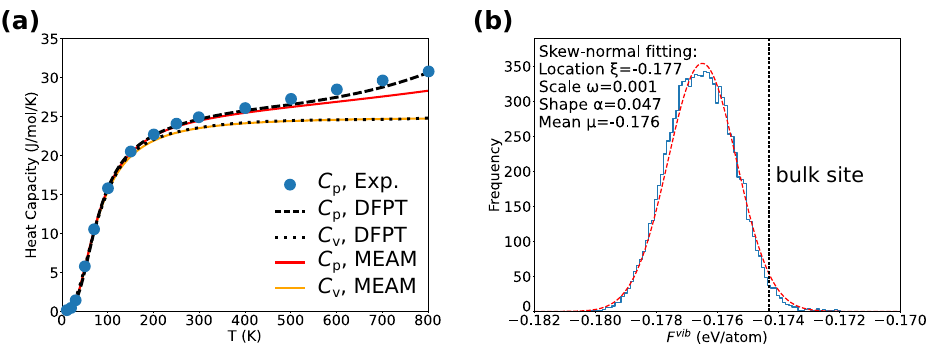}
\caption{(a) The vibrational heat capacities at constant volume $C_{\text{v}}$ and pressure $C_{\text{p}}$ of bulk Mg (containing 5 $\times$ 5 $\times$ 5 orthogonal unit cells) as functions of temperature, calculated by the full harmonic approximation (HA) method using the MEAM potential. For comparison, results from density functional perturbation theory (DFPT)~\cite{nie2007ab} and experimental data~\cite{hultgren1973selected} are presented. (b) The distribution of $F^{\text{vib}}$ of all GB sites in the Mg polycrystal at 600 K calculated based on HA ($r_{\text{HA}}$=14 $\text{\AA}$, bin size: 1$\times10^{-4}$ eV/atom). A skew-normal function (dashed curve) was fitted to the $F^{\text{vib}}$ distribution. The dashed line indicates the $F^{\text{vib}}$ value for bulk Mg.
}
\label{fig1}
\end{figure*}

To benchmark the performance of the MEAM potential in predicting thermal properties of pure Mg, the temperature-dependent vibrational heat capacities at constant volume $C_{\text{v}}$ and pressure $C_{\text{p}}$ were calculated (see Section~\ref{methods}) and compared with ab-initio~\cite{nie2007ab} and experimental~\cite{hultgren1973selected} data, see Figure~\ref{fig1}(a). The calculated temperature-dependent $C_{\text{v}}$ using the MEAM potential shows excellent agreement with the ab-initio result. In addition, the calculated $C_{\text{p}}$ using the MEAM potential aligns closely with the ab-initio data and experimental measurements. Overall, these comparisons demonstrate that the MEAM potential effectively describes the thermal properties of pure Mg.

The per-site vibrational free energies $F^{\text{vib}}$ for over 38,000 GB sites within the Mg polycrystal, as characterized using the adaptive CNA method, were calculated following Section~\ref{methods}. The distribution of $F^{\text{vib}}$ at 600 K across these GB sites is shown in Figure~\ref{fig1}(b), fitted with a skew-normal function characterized by the following parameters: location $\xi$= $-$0.177 eV/atom, scale $\omega$= 0.001 eV/atom, and shape $\alpha$= 0.047, leading to a mean value $\mu$ of $-$0.176 eV/atom. The small $\alpha$ value alongside the closely matched $\xi$ and $\mu$ values suggests that the distribution is nearly symmetric, closely approximating a normal distribution. Notably, a clear deviation from the $F^{\text{vib}}$ value of the bulk site at 600 K ($-$0.1743 eV/atom) was identified, highlighting significant differences in thermal properties between GB and bulk sites.

\subsection{Segregation free energy at finite temperatures}

\begin{figure*}[hbt!]
\centering
\includegraphics[width=\textwidth]{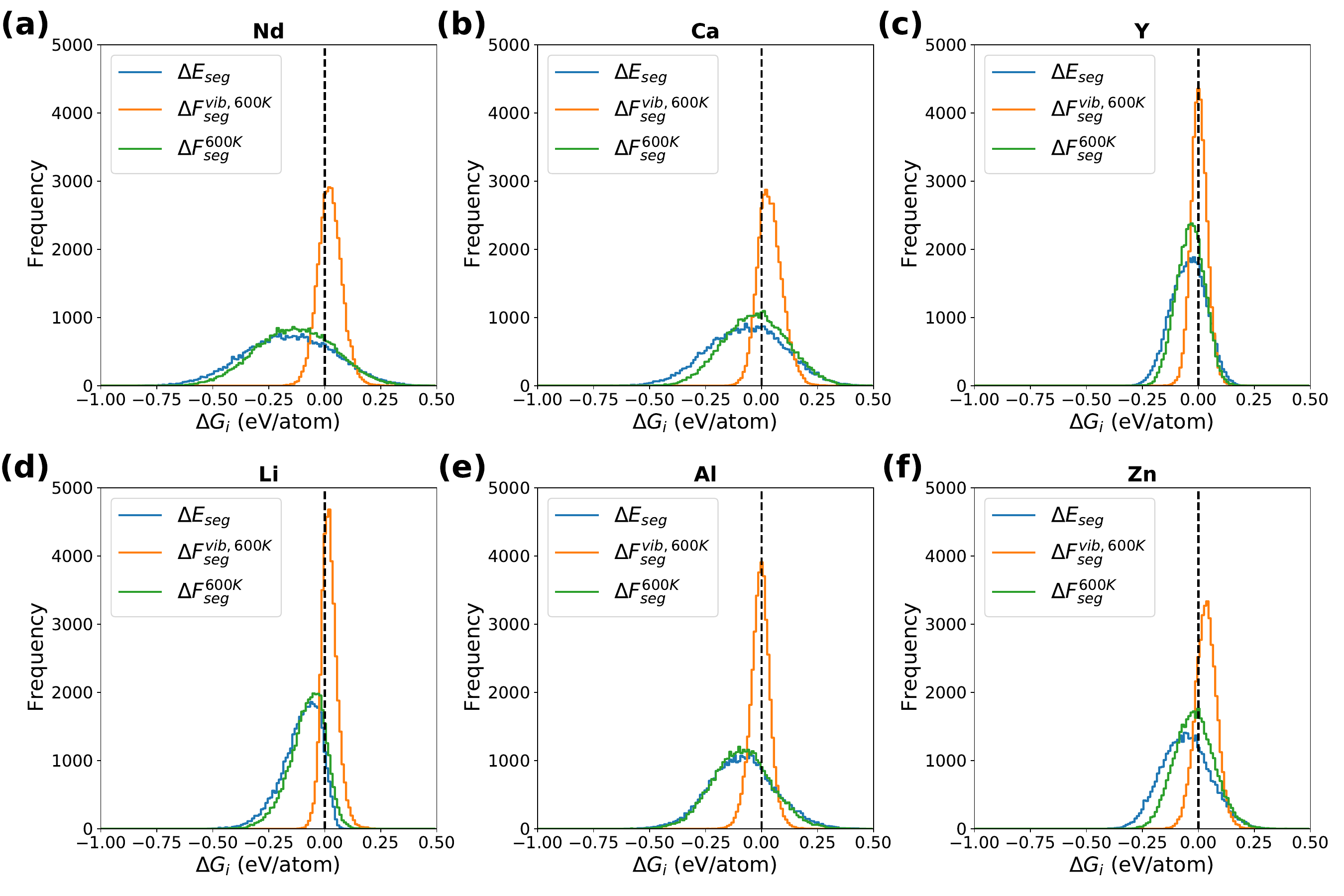}
\caption{Distributions of segregation energy $\Delta E_{\text{seg}}$ at 0 K, segregation vibrational free energy $\Delta F_{\text{seg}}^{\text{vib}}$ and segregation free energy $\Delta F_{\text{seg}}$ at 600 K of (a) Nd, (b) Ca, (c) Y, (d) Li, (e) Al, and (f) Zn solutes at GB sites in the Mg polycrystal using atomistic simulations. Bin size: 0.01 eV/atom.}
\label{fig2}
\end{figure*}

\begin{table*}[!htbp]
\centering
\caption[]{\label{tab1}Summary of fitting parameters, including location ($\xi$, in eV/atom), scale ($\omega$, in eV/atom), and shape ($\alpha$, dimensionless) parameters, as well as mean value ($\mu = \xi + \omega \frac{\alpha}{\sqrt{1+\alpha^2}} \sqrt{\frac{2}{\pi}}$, in eV/atom) for skew-normal distribution applied to segregation energy $\Delta E_{\text{seg}}$ at 0 K, segregation vibrational free energy $\Delta F_{\text{seg}}^{\text{vib}}$ and segregation free energy $\Delta F_{\text{seg}}$ at 600 K in Mg-X (X=Nd, Ca, Y, Li, Al and Zn) alloy systems.}
\centering
\scriptsize
\begin{tabular}{p{0.15\textwidth}p{0.06\textwidth}p{0.06\textwidth}p{0.06\textwidth}p{0.06\textwidth}p{0.06\textwidth}p{0.06\textwidth}p{0.06\textwidth}p{0.06\textwidth}p{0.06\textwidth}p{0.06\textwidth}p{0.06\textwidth}p{0.06\textwidth}}
\hline\hline
\addlinespace[0.1cm]
\multicolumn{1}{l}{} & \multicolumn{4}{c}{$\Delta E_{\text{seg}}$ at 0 K} & \multicolumn{4}{c}{$\Delta F_{\text{seg}}^{\text{vib}}$ at 600 K} & \multicolumn{4}{c}{$\Delta F_{\text{seg}}$ at 600 K}\\
\cmidrule(lr){2-5}
\cmidrule(lr){6-9}
\cmidrule(lr){10-13}
Solute element & $\xi$ & $\omega$ & $\alpha$ & $\mu$ & $\xi$ & $\omega$ & $\alpha$ & $\mu$ & $\xi$ & $\omega$ & $\alpha$ & $\mu$ \\
\addlinespace[0.1cm]
\hline
\addlinespace[0.1cm]
Nd & $-$0.078 & 0.212 & $-$0.507 & $-$0.154 & $-$0.009 & 0.065 & 0.990 & 0.027 & $-$0.077 & 0.180 & $-$0.368 & $-$0.127 \\
Ca & 0.019 & 0.188 & $-$0.669 & $-$0.064 & $-$0.007 & 0.074 & 1.568 & 0.043 & $-$0.074 & 0.151 & 0.488 & $-$0.021 \\
Y & 0.006 & 0.089 & $-$0.810 & $-$0.039 & $-$0.016 & 0.044 & 1.006 & 0.009 & $-$0.000 & 0.069 & $-$0.651 & $-$0.030 \\
Li & 0.009 & 0.140 & $-$4.527 & $-$0.100 & $-$0.009 & 0.051 & 1.851 & 0.027 & 0.016 & 0.120 & $-$2.584 & $-$0.074 \\
Al & $-$0.169 & 0.170 & 0.853 & $-$0.081 & 0.025 & 0.051 & $-$0.801 & $-$0.001 & $-$0.157 & 0.152 & 0.793 & $-$0.081 \\
Zn & $-$0.124 & 0.130 & 0.975 & $-$0.051 & 0.016 & 0.053 & 0.628 & 0.039 & $-$0.077 & 0.110 & 1.092 & $-$0.012 \\
\addlinespace[0.1cm]
\hline\hline
\end{tabular}
\end{table*}

The per-site segregation energies $\Delta E_{\text{seg}}$ for solute elements (Nd, Ca, Y, Li, Al, and Zn) at 0 K of GB sites within the Mg polycrystal were calculated. The elements Nd, Ca, and Y are considered larger solutes relative to the Mg atom, as substituting Mg with these elements induces positive volumetric strain in the Mg lattice. In contrast, the elements Li, Al, and Zn, having smaller atomic radii, introduce negative volumetric strain to the Mg matrix upon substitution~\cite{wang2024defects}. The resulting $\Delta E_{\text{seg}}$ distributions in these systems are illustrated in Figure~\ref{fig2}. For all simulated Mg alloys in this work, the $\Delta E_{\text{seg}}$ spectra exhibit a skew-normal distribution, similar to observations in FCC polycrystalline systems as reported by~\cite{wagih2019spectrum,wagih2020learning}. A skew-normal function was applied to fit the $\Delta E_{\text{seg}}$ spectra, with the fitting parameters $\xi$, $\omega$ and $\alpha$ detailed in Table~\ref{tab1}. 
The majority of the GB sites across all studied Mg alloy systems are energetically favorable to solute segregation, as evidenced by their negative $\Delta E_{\text{seg}}$. This trend is further underscored by the mean ($\mu$) values of the $\Delta E_{\text{seg}}$ spectra, which are also negative for all systems. Notably, Nd shows the most negative $\mu$ value of $-$0.154 eV/atom, with Li and Al following with $\mu$ of $-$0.100 and $-$0.081 eV/atom, respectively, highlighting their strong segregation tendencies.

The skewness parameter ($\alpha$) reflects the asymmetry in segregation energy distributions. For the Nd, Ca, Y, and Li solutes, the distribution of $\Delta E_{\text{seg}}$ exhibits a left-skewed pattern with the negative mean being pulled toward the left tail, as evidenced by negative $\alpha$ values. This indicates a strongly favorable segregation tendency, as both the central tendency and extreme tails favor segregation. On the other hand, the Al and Zn solutes demonstrate a right-skew distribution, characterized by positive $\alpha$ values, indicating that the negative mean of these distributions is shifted towards the right tail. This suggests mixed segregation behavior, where most sites are favorable, but a significant number of less favorable sites exist. Among the solute elements studied, Nd exhibits the broadest distribution of $\Delta E_{\text{seg}}$ followed by Ca, as indicated by the values of $\omega$, which suggests greater variability in segregation behavior across GB sites.

\begin{figure*}[hbt!]
\centering
\includegraphics[width=\textwidth]{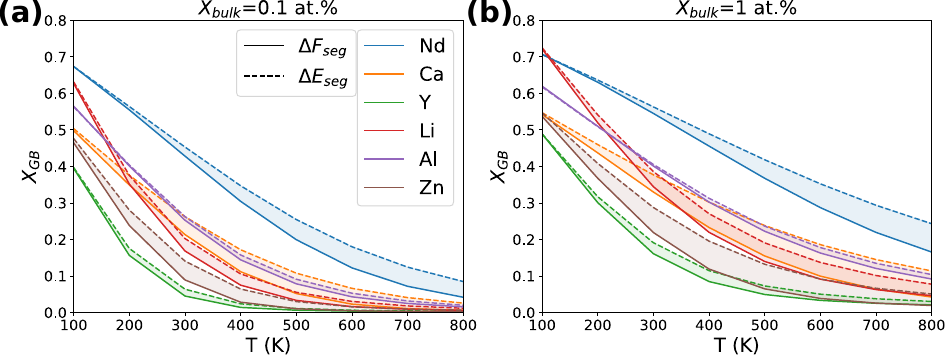}
\caption{Grain boundary concentration of X solutes (X=Nd, Ca, Y, Li, Al, and Zn) as a function of temperature with and without considering the vibrational entropy contribution at (a) 0.1 at.\% and (b) 1 at.\% bulk solute concentrations following the Langmuir–McLean model, as denoted in Equations~\eqref{equ11} and~\eqref{equ12}. The shaded areas indicate the differences between predicted GB solute concentrations with and without considering the vibrational entropy contribution.}
\label{fig3}
\end{figure*}

The segregation vibrational free energies $\Delta F_{\text{seg}}^{\text{vib}}$ at finite temperatures were calculated by the HA approach according to Section~\ref{methods}. The distribution of $\Delta F_{\text{seg}}^{\text{vib}}$ follow a skew-normal function in all simulated Mg alloy systems, similar to the $\Delta E_{\text{seg}}$ distributions. The spread of the $\Delta F_{\text{seg}}^{\text{vib}}$ distributions at 600 K is significantly narrower than that of $\Delta E_{\text{seg}}$, with $\omega$ values spanning from 0.044 to 0.074 eV/atom. Notably, Y not only presents the narrowest distribution within the $\Delta E_{\text{seg}}$ spectrum but also exhibits the most constrained spread in the $\Delta F_{\text{seg}}^{\text{vib}}$ spectrum. The $\mu$ values are slightly positive for all alloy systems except for Al, which exhibits a $\mu$ value of $-$0.001 eV/atom. In addition, Al is distinguished as the only element exhibiting a negative $\alpha$ value of $-$0.801 in the $\Delta F_{\text{seg}}^{\text{vib}}$ spectra, demonstrating a left-skewed distribution where the mean is pulled towards the left tail.

The segregation free energies $\Delta F_{\text{seg}}$ at finite temperatures were determined by adding the vibrational entropy contribution to the segregation energy $\Delta E_{\text{seg}}$ (Section~\ref{methods}). The $\Delta F_{\text{seg}}$ distributions of the Mg alloy systems were fitted using a skew-normal function as shown in Figure~\ref{fig2}. Compared to the $\Delta E_{\text{seg}}$ distributions, a noticeable shift towards positive values was observed for all solute elements, implying a reduced tendency for segregation as the temperature rises. Despite this shift, all $\mu$ values of the $\Delta F_{\text{seg}}$ distribution remained negative at 600 K, with Nd showing the most negative value at $-$0.127 eV/atom. For Al, the transition in segregation behavior with temperature is less marked (see Figure~\ref{fig2}(e)), as the $\mu$ value of the $\Delta F_{\text{seg}}$ distribution remains consistent with that of the $\Delta E_{\text{seg}}$ distribution, both measured at $-$0.081 eV/atom. A contraction in the distributions of $\Delta F_{\text{seg}}$ was evident from the decrease in the $\omega$ value, suggesting reduced variability in segregation tendencies at finite temperatures. Notably, the distribution for Ca demonstrated a significant change in skewness at 600 K, with the skewness parameter $\alpha$ shifting from $-$0.669 at 0 K to a positive value of 0.488. This shift in $\alpha$ indicates a transition in segregation behavior, where favorable segregation sites become less dominant, reflecting a more balanced distribution of segregation tendencies across the GB sites.

Besides quantifying energy spectra, the per-site segregation (free) energy, denoted as $\Delta G_i$, can be directly correlated with solute concentration at GBs based on the Langmuir–McLean model, as illustrated in Equation~\eqref{equ11} in the context of an infinitesimally dilute solid solution: 
\begin{equation}
\label{equ11}
X_{\text{GB},i} = \left(1 + \frac{1 - X_{\text{bulk}}}{X_{\text{bulk}}} \exp\left(\frac{\Delta G_i}{k_\text{B} T}\right)\right)^{-1},
\end{equation}
where $X_{\text{bulk}}$ is the solute concentration in the bulk region, and $X_{\text{GB},i}$ is the fraction of solute concentration of a GB site $i$. In this work, $\Delta G_i$ is referred to as the Gibbs free energy change, which is traditionally used in the Langmuir-McLean model and is approximately equivalent to the Helmholtz free energy change under the simulation conditions, as the pressure-volume work term is negligible in the context of bulk solids. 
This correlation facilitates the calculation of the weighted average solute concentration at GBs ($X_\text{GB}$) as described in Equation~\eqref{equ12}: 
\begin{equation}
\label{equ12}
X_\text{GB} = \sum_i F_i X_{\text{GB},i},
\end{equation}
where $F_i$ is the fraction of GB sites with the solute concentration $X_{\text{GB},i}$. Such a methodological approach allows for direct comparison and correlation of theoretical predictions from atomic-scale modeling with experimentally measured solute concentrations at GBs.

The effect of vibrational entropy on solute segregation at GBs was illustrated by calculating the fractional solute concentration of individual sites according to Equation~\eqref{equ11} with and without considering the vibrational term $\Delta F_{\text{seg}}^{\text{vib}}$. The resulting concentrations of various solutes (Nd, Ca, Y, Li, Al, and Zn) at GBs within the simulated polycrystalline Mg at $X_{\text{bulk}}$ of 0.1 at.\% and 1 at.\% as a function of temperature are shown in Figure~\ref{fig3}.
Accounting for the vibrational entropy contribution results in notably lower solute concentrations at GBs than those derived from calculations relying solely on $\Delta E_{\text{seg}}$. The difference in solute concentrations at GBs, attributed to vibrational entropy, becomes increasingly significant with rising temperature and bulk concentration $X_{\text{bulk}}$. Among the solutes studied, Nd stands out for demonstrating a significantly higher $X_\text{GB}$ value across the temperature range up to 800 K examined. This selected temperature range slightly exceeds the conventional upper limit of thermomechanical processing temperatures for Mg alloys in experimental settings, which typically extend up to 723 K. This choice ensures that the temperature range of interest is fully covered while allowing for insights into segregation behavior under elevated temperature conditions.

\subsection{Machine learning prediction of segregation energy and segregation free energy}

\begin{figure*}[hbt!]
\centering
\includegraphics[width=\textwidth]{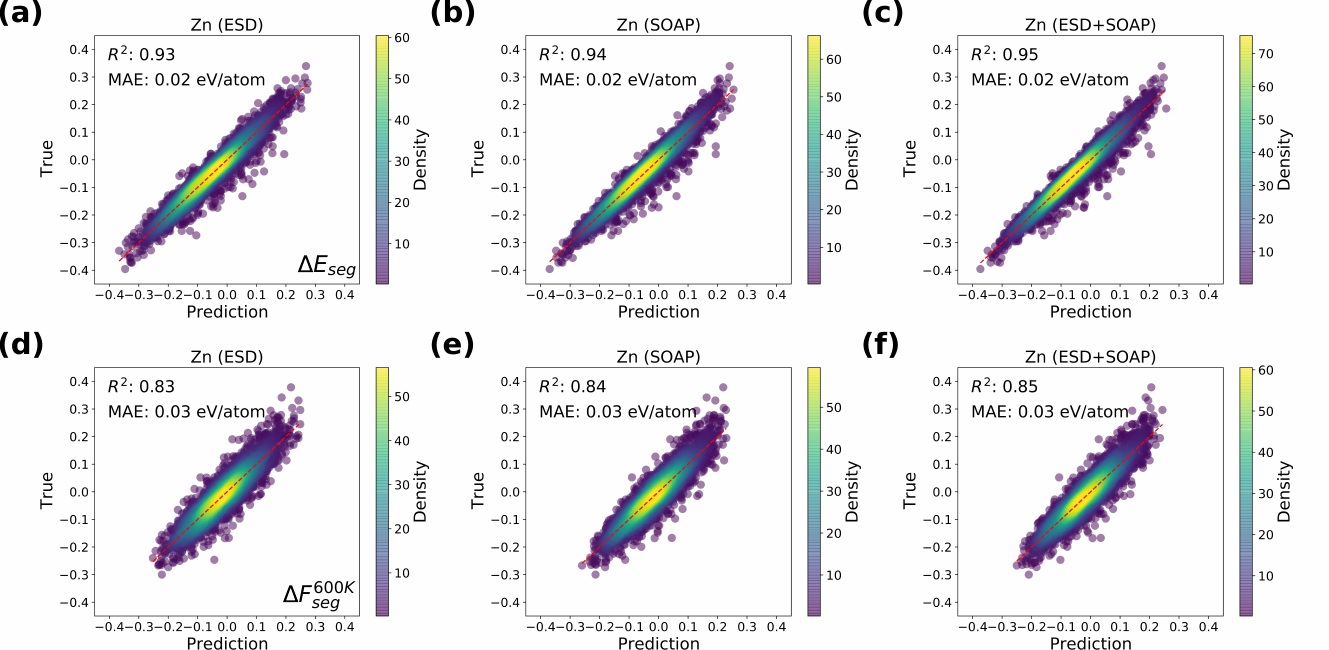}
\caption{Contour plots of ML predicted versus true (a-c) per-site segregation energy $\Delta E_{\text{seg}}$ and (d-f) segregation free energy $\Delta F_{\text{seg}}$ at 600 K of Zn solute segregation at GB sites in the Mg polycrystal. Only the validation data are illustrated here. The CatBoostRegressor model was trained separately using the (a,d) ESD features, (b,e) SOAP features, and (c,f) both ESD and SOAP features. The color density indicates the concentration of data points, with denser regions shown in lighter shades. The diagonal red dashed line represents the ideal scenario where predicted values match the true values perfectly.}
\label{fig4}
\end{figure*}

Estimating the solute segregation tendency, particularly calculating the vibrational entropy contribution via the HA approach, is computationally expensive for all GB sites in polycrystals. The ML approach is ideal for developing a model to predict the segregation tendency of solutes to GB sites and capture the segregation energy spectrum based on accessible energetic and structural features, which are computationally inexpensive. To predict $\Delta E_{\text{seg}}$, $\Delta F_{\text{seg}}^{\text{vib}}$, and $\Delta F_{\text{seg}}$ at 600 K, we generated 16 ESD that capture the local atomic environment and thermal vibrational properties of GB sites in the polycrystal. The Pearson's correlation matrices of these features with target segregation energies in different Mg alloy systems are illustrated in Figures S4-9 (Supplementary Material) to identify which features have strong relationships with the target variables and with each other. Among all ESD, hydrostatic stress $\sigma_{h}$, which is calculated based on $\sigma_{xx}$, $\sigma_{yy}$, and $\sigma_{zz}$, shows the highest correlation coefficient above 0.80 with $\Delta E_{\text{seg}}$ and $\Delta F_{\text{seg}}$ at 600 K for all solute elements, indicating its significant role in determining the GB segregation behavior of solute atoms in the material. The normal stress components $\sigma_{xx}$, $\sigma_{yy}$, and $\sigma_{zz}$ exhibit similar correlation coefficients. Additionally, potential energy, atomic volume, surface area, and coordination number also exhibit high correlation coefficients with the segregation energies. 

\begin{figure*}[hbt!]
\centering
\includegraphics[width=\textwidth]{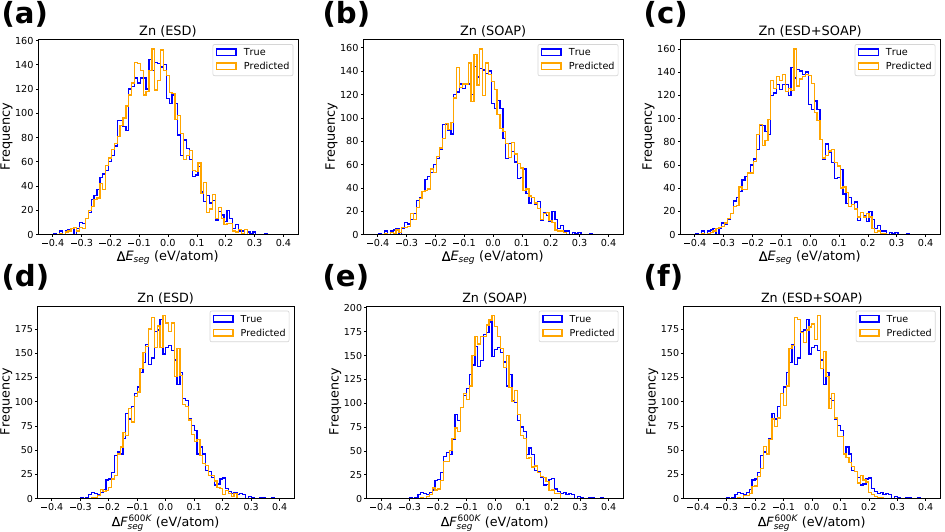}
\caption{Distributions of ML predicted and true (a-c) per-site segregation energy $\Delta E_{\text{seg}}$ and (d-f) segregation free energy $\Delta F_{\text{seg}}$ at 600 K of Zn solute segregation at GB sites in the Mg polycrystal. Only the validation data are illustrated here. The CatBoostRegressor model was trained separately using the (a,d) ESD features, (b,e) SOAP features, and (c,f) both ESD and SOAP features. Distributions of predicted versus true energies with a bin size of 0.01 eV/atom.}
\label{fig5}
\end{figure*}

The contour plots in Figure~\ref{fig4}(a) reveal that the predicted $\Delta E_{\text{seg}}$ values align closely with the true values along the diagonal red dashed line, which represents the ideal prediction scenario. The color density represents the concentration of data points, with denser regions displayed in brighter shades. This emphasizes that the highest concentration of data points is near the ideal prediction line, indicating the strong predictive capability of the CatBoostRegressor model with the ESD feature set.
The good performance of the CatBoostRegressor model in predicting $\Delta E_{\text{seg}}$ with the ESD features is further demonstrated by the $R^{2}$ scores, with $R^{2}$ between 0.82 and 0.87 for Nd, Y and Al, and between 0.90 and 0.94 for Ca, Li, and Zn, see Figure~\ref{fig4} and Figure S10. In addition, the distributions of the ESD exhibit a high degree of overlap between the predicted and the calculated $\Delta E_{\text{seg}}$, as shown in Figure~\ref{fig5} and Figure S11 for all solute elements. Among all ESD features, $\sigma_{h}$ demonstrates the highest importance, except for Li solute, where the potential energy was identified as the most important feature, see Figure S12. Potential energy was ranked as the second most important feature for Al and Zn solute elements and the third for Ca. The structural features obtained using Voronoi tessellation, including atomic volume, surface area, and cavity radius, were ranked highly in importance, e.g., for Nd and Zn solute elements, atomic volume is the second most important feature, while for Y, cavity radius was ranked second. The vibrational properties calculated from the series of MD simulations, including vibrational MSD $MSD_{\text{vib}}$ and flexibility volume $V_{\text{flex}}$, were demonstrated as important features in predicting $\Delta E_{\text{seg}}$: for Nd, Ca, Y, Al, and Zn, these vibrational features are among the top five most important features.

The performance of the ML approach in predicting $\Delta F_{\text{seg}}^{\text{vib}}$ at 600 K is much lower compared to $\Delta E_{\text{seg}}$, as demonstrated by the scattered contour plots of predicted versus true $\Delta F_{\text{seg}}^{\text{vib}}$ values (Figure S13) and the not well-matched distributions (Figure S14). Among all solute elements, Ca shows the highest $R^{2}$ score of 0.52, followed by Zn at 0.45 and Y at 0.44. Additionally, the prediction of the distributions of $\Delta F_{\text{seg}}^{\text{vib}}$ is less satisfactory, especially for the tails of the spectra where the GB sites exhibit high or low $\Delta F_{\text{seg}}^{\text{vib}}$ values, see Figure S14. Among all features, $\Delta E_{\text{seg}}$ is the most important, while other ESD features exhibit much lower importance factors (Figure S15). For Nd and Li, the vibrational free energy $F^{\text{vib}}$ at 600 K of the pure Mg counterpart demonstrates the second highest importance, while for the rest, potential energy takes second place.

For the prediction of $\Delta F_{\text{seg}}$ at 600 K, the ESD-featured CatBoostRegressor model shows relatively good performance, with all solute elements exhibiting $R^{2}$ scores above 0.75, see Figure~\ref{fig4}(d) and Figure S16. The Ca solute element shows the highest $R^{2}$ score of 0.86, followed by Zn at 0.83. Additionally, the distributions of the predicted $\Delta F_{\text{seg}}$ generally show good agreement with the true spectra (Figure~\ref{fig5}). However, for the Y solute element, which exhibits the lowest performance with a $R^{2}$ score of 0.76, the tails of the spectrum where the GB sites exhibit high or low $\Delta F_{\text{seg}}$ were underestimated, see Figure S17. The ranking of feature importance for the $\Delta F_{\text{seg}}$ prediction is similar to that of the $\Delta E_{\text{seg}}$ prediction, with $\sigma_{h}$, potential energy and the Voronoi-based structural features showing high importance (Figure S18). 

\begin{figure*}[hbt!]
\centering
\includegraphics[width=\textwidth]{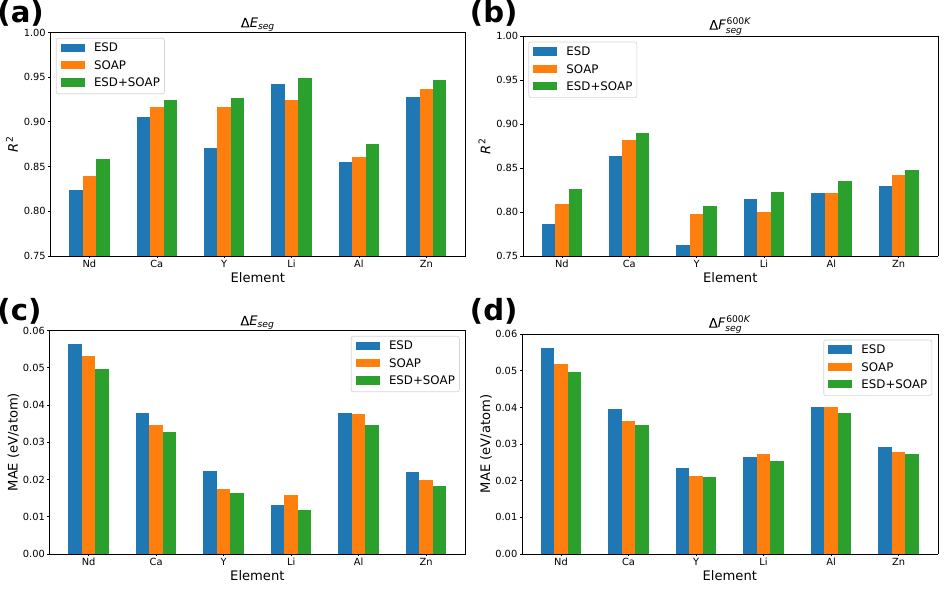}
\caption{Performance of the CatBoostRegressor model in predicting (a,c) per-site segregation energy $\Delta E_{\text{seg}}$ and (b,d) segregation free energy $\Delta F_{\text{seg}}$ at 600 K at GB sites belong to the validation data in the Mg polycrystal using different feature sets for various elements. The model's accuracy and error were evaluated based on (a,b) $R^{2}$ score and (c,d) mean absolute error (MAE).} 
\label{fig6}
\end{figure*}

In addition to the ESD features, we generated SOAP features for automatically describing the fingerprint of the local atomic environment at GB sites in the polycrystal. These SOAP descriptors were fed to the CatBoostRegressor model to predict $\Delta E_{\text{seg}}$, $\Delta F_{\text{seg}}^{\text{vib}}$, and $\Delta F_{\text{seg}}$ at 600 K (Figure~\ref{fig4}-\ref{fig5} and Figures S19-24). The SOAP-featured model shows better performance in predicting $\Delta E_{\text{seg}}$ and $\Delta F_{\text{seg}}$ at 600 K than the ESD-featured model for all solute elements, except for Li, as demonstrated by higher $R^{2}$ scores and lower MAE (see Figure~\ref{fig6}). However, the performance in predicting $\Delta F_{\text{seg}}^{\text{vib}}$ remains poor (Figures S21,22). By merging ESD and SOAP descriptors, the ML model shows further improvement in predicting $\Delta E_{\text{seg}}$ and $\Delta F_{\text{seg}}$, as evidenced by the tighter clustering around the diagonal (Figure~\ref{fig4}(c,f)) and with the $R^{2}$ score of all solute element in predicting $\Delta E_{\text{seg}}$ above 0.85 and $\Delta F_{\text{seg}}$ above 0.80 (Figures S25-30). However, the improvement by utilizing combined ESD and SOAP features in predicting $\Delta F_{\text{seg}}^{\text{vib}}$ is limited. The model's performance in predicting $\Delta F_{\text{seg}}$, which reflects a rule of mixture, combining the high accuracy of $\Delta E_{\text{seg}}$ predictions with the relatively lower accuracy of $\Delta F_{\text{seg}}^{\text{vib}}$ predictions. This behavior arises because $\Delta F_{\text{seg}}$ is a linear combination of $\Delta E_{\text{seg}}$ and $\Delta F_{\text{seg}}^{\text{vib}}$ as illustrated in Equation~\eqref{equ3}.

The potential redundancy and overfitting in the training features were evaluated. For instance, some of the ESDs may contain overlapping information, such as atomic volume being related to both hydrostatic stress and flexibility volume. However, the performance metrics of the model across training, testing, and unseen validation datasets indicate only slight differences, which suggests that overfitting is not occurring, as demonstrated in Table S2.
Additionally, a learning curve analysis by plotting the training and validation loss against the size of the training dataset shows an initial low training error, which gradually increases and then stabilizes as the training size grows, see Figure S31. This behavior indicates that the model tends to memorize patterns with smaller datasets, resulting in low training error. However, as more data is introduced, the model shifts from memorization to generalization, leading to a rise in training error. Eventually, the training error plateaus when the model reaches its learning capacity, meaning additional data does not significantly impact performance. The comparable validation errors further confirm that the model is well-suited for this study and does not suffer from overfitting. Furthermore, the feature overlap does not appear to significantly impact predictive accuracy as the effect of dimensionality reduction by removing highly correlated ESD features (correlation threshold = 0.9) did not result in any significant improvement in model performance, with both MAE and $R^{2}$ values remaining nearly identical, see Table S3.

For the SOAP features, dimensionality reduction using Principal Component Analysis (PCA) or feature selection techniques by selecting the top-20 important SOAP features based on feature importance rankings were applied. As shown in Table S4, neither using 20 PCA components nor restricting the model to the top-20 important features resulted in improved performance compared to the full set of SOAP features. The MAE and $R^{2}$ values for these reduced models indicate a slight decrease in prediction accuracy. While dimensionality reduction did not enhance accuracy, it remains a valuable tool for improving the model's interpretability, particularly in scenarios where computational resources or explainability are key considerations.

The residual distributions of predicted and true $\Delta E_{\text{seg}}$, $\Delta F_{\text{seg}}^{\text{vib}}$, and $\Delta F_{\text{seg}}$ via ESD, SOAP, and combined ESD and SOAP features are shown in Figures S32-34. A left-skewed residual distribution in $\Delta E_{\text{seg}}$ and $\Delta F_{\text{seg}}$ was observed for all solute elements and feature sets, indicating the models consistently over or under-predict the solute segregation tendency at favorable segregation sites. For future work, uncertainty determination will be crucial to better understand and address these biases and refine the models accordingly.

\section{Discussion}

\subsection{Signature energetic and structural descriptors}

The SOAP-featured ML model demonstrated superior performance in predicting $\Delta E_{\text{seg}}$ and $\Delta F_{\text{seg}}$ at 600~K at GB sites in the Mg polycrystal compared to the ESD-featured model. The SOAP descriptors define the Gaussian smeared atomic density constructed on each site, effectively representing the local atomic environment~\cite{bartok2010gaussian,bartok2013representing}. While SOAP has been widely utilized for generating structural features to train ML models for predicting not only $\Delta E_{\text{seg}}$ but also other GB properties~\cite{homer2022examination,fujii2020quantitative}, it has certain limitations. These include the statistical significance of using a large number of features (over a thousand variables) in the training procedure, which raises the risk of learning spurious correlations, and the absence of physical insights obtained from this approach. In contrast, the ESD feature set comprises important properties of the local atomic environment, each with its own physical significance. This can aid in gaining a deeper understanding of the dominant factors influencing segregation behavior and in developing physics-based ML models. 

Among all ESD features, $\sigma_{h}$ stands out as the most important feature in predicting $\Delta E_{\text{seg}}$ and $\Delta F_{\text{seg}}$. For solute elements with larger atomic radii than Mg, a negative correlation with target segregation energies was observed, while for solute elements with smaller atomic radii, a positive correlation was identified (see Figures S4-9). Such solute size-dependent segregation behavior has been widely reported in previous experimental and atomic-scale modeling works~\cite{nie2013periodic,huber2014atomistic,he2021unusual,xie2021nonsymmetrical}, namely, the preference for large solutes to segregate at sites with extensive strain and small solutes to segregate at sites with compressive strain. Huber et al.~\cite{huber2014atomistic} developed a linear elastic model for predicting $\Delta E_{\text{seg}}$ based on this concept, where the multiplication of variate in atomic volume and bulk modulus estimates the isotropic stress state at the GB site.

Among all studied solute elements, the prediction of $\Delta E_{\text{seg}}$ for Li shows the best performance. Additionally, Li is the only solute element that exhibits better performance with the ESD features than the SOAP descriptors. Furthermore, unlike other solute elements for which $\sigma_{h}$ stands out as the most important feature, potential energy is the most important ESD feature in predicting $\Delta E_{\text{seg}}$ and $\Delta F_{\text{seg}}$ for Li (see Figure S12). These could be attributed to the size-dependent segregation behavior, where Li exhibits the closest atomic radius to Mg among all studied solute elements in terms of the variate in atomic volume after introducing the substitutional solute in the Mg matrix~\cite{wang2024defects}. Due to the similar atomic radii, Li is expected to introduce minimal lattice distortion or have a lesser effect on minimizing the strain energy. Therefore, the structural features and virial stress tensors that encapsulate the gradient of the potential energy and atomic volume become less relevant in describing its segregation tendency.

Apart from the common ESD features considered in previous ML approaches for predicting $\Delta E_{\text{seg}}$~\cite{huber2018machine,messina2021machine,mahmood2022atomistic,ma2023efficient,borges2024insights}, such as atomic volume and stress tensors, this study included vibrational properties ($MSD_{\text{vib}}$ and $V_{\text{flex}}$) in the ESD feature set, which demonstrated high importance (Figure S12). Flexibility volume ($V_{\text{flex}}$) is a structural indicator designed to enhance the understanding and prediction of metallic glass properties~\cite{ding2016universal}. It is defined as a volume-scaled (or density-normalized) vibrational MSD ($MSD_{\text{vib}}$). The underlying physics principle of $V_{\text{flex}}$ is rooted in its connection to the Debye model. By normalizing $MSD_{\text{vib}}$ with the average atomic spacing, $V_{\text{flex}}$ provides a dimensionless measure that scales with free volume while incorporating dynamic information. Unlike traditional indicators such as free volume, $V_{\text{flex}}$ combines static structural data with dynamic properties, offering a more comprehensive assessment of the local atomic environment and the degree of atomic vibrational freedom. This makes $V_{\text{flex}}$ an ideal feature for predicting "potential" excess free volume~\cite{pei2023atomistic} in addition to the excess free volume at individual GB sites. As demonstrated in our previous investigation on solute segregation at general GBs, the linear elastic model fails to accurately predict $\Delta E_{\text{seg}}$ as the structural relaxation after introducing a solute atom involves atomic rearrangement beyond the nearest neighbors~\cite{pei2023atomistic}. Therefore, if we consider the excess free volume characterized by atomic volume as the principal factor, the flexibility of the local atomic environment described by $V_{\text{flex}}$ could contribute to the segregation tendency in a second-order manner.

Although the prediction of overall free energy of solute segregation is reliable, the performance of the ML approach in predicting the vibrational entropy contribution $\Delta F_{\text{seg}}^{\text{vib}}$ is unsatisfactory using both ESD and SOAP feature sets. $\Delta F_{\text{seg}}^{\text{vib}}$ is directly related to segregation entropy ($\Delta S_{\text{seg}}$), as demonstrated in Equation~\eqref{equ2} and~\eqref{equ3}. Rittner and Seidman~\cite{rittner1997solute} reported a strong linear correlation between $\Delta E_{\text{seg}}$ and $\Delta S_{\text{seg}}$ for Pd solute segregation in symmetric tilt Ni GBs using atomistic simulations. The experimental data on Si, P, and C solute segregation in tilt GBs in Fe-Si alloy systems from Lejček and Hofmann~\cite{lejvcek1995thermodynamics} also demonstrated a linear segregation energy-entropy relationship. Tuchinda and Schuh~\cite{tuchinda2023vibrational} reported the linear segregation energy-entropy relationship in the Ni-based polycrystalline alloys using atomistic simulations and attributed it to the physical principle that deeper energy wells are narrower, thus restricting vibration. However, significant scattering was detected and fitted to a bivariate spectral isotherm~\cite{tuchinda2023vibrational}. Similar bivariate skew-normal distributions were also identified in the Mg alloying systems in this study, as illustrated in Figure S35. A linear function was fitted to the distribution, with the slope having a similar magnitude ($10^{-4}$ 1/K) as that obtained for the Ni-based alloy systems~\cite{tuchinda2023vibrational}. 
Such significant scattering between $\Delta E_{\text{seg}}$ and $\Delta F_{\text{seg}}^{\text{vib}}$ leads to the unsatisfactory prediction of $\Delta F_{\text{seg}}^{\text{vib}}$ using the $\Delta E_{\text{seg}}$ feature. The prediction becomes even worse without including $\Delta E_{\text{seg}}$ in the feature sets, as $\Delta E_{\text{seg}}$ exhibits the highest correlation with $\Delta F_{\text{seg}}^{\text{vib}}$ among all features, as illustrated in the correlation matrices (Figures S4-9) and feature importance (Figure S15).
The SOAP feature set, which is believed to comprehensively capture the structural features of the local atomic environment, fails to predict $\Delta F_{\text{seg}}^{\text{vib}}$, implying that structural features before solute segregation alone are insufficient for predicting vibrational entropy contribution of solute segregation. This could be interpreted as the change in bond stiffness, crucial in determining the variation of vibrational frequency when introducing a solute at a GB site, which cannot be precisely described using the structural features of the pre-segregation local atomic environment. Other energetic features describing the system before solute segregation, such as potential energy and $F^{\text{vib}}$, although still among the highest important features, exhibit much lower importance compared to $\Delta E_{\text{seg}}$ (Figure S15). Overall, these outcomes highlight that the features after segregation could be crucial in predicting the vibrational entropy contribution, with $\Delta E_{\text{seg}}$ being particularly important as it describes the energy difference before and after introducing the solute element. For better prediction of $\Delta F_{\text{seg}}^{\text{vib}}$, incorporating more structural and energetic features after solute segregation would be beneficial.

Despite the relatively low prediction accuracy of per-site $\Delta F_{\text{seg}}^{\text{vib}}$, the overall trend of its contribution to segregation behavior, specifically, the averaged solute concentration at GBs as calculated using the Langmuir-McLean model, is well captured. As illustrated in Figure S36, the solute concentration at 600 K, when computed using the predicted $\Delta F_{\text{seg}}$ values, deviates only slightly from the values obtained using the true $\Delta F_{\text{seg}}$ values (see Figure S36(b)). The ML model's performance in predicting solute concentration at finite temperatures is comparable to its performance in predicting solute concentration at 0 K using $\Delta E_{\text{seg}}$ alone (see Figure S36(a)). This observation aligns with the relatively small absolute value of $\Delta F_{\text{seg}}^{\text{vib}}$ in comparison to segregation energy, reinforcing its limited impact on segregation trends. These results demonstrate the capability of the ML model in predicting solute concentration at finite temperatures, which serves as a crucial parameter for comparison with experimental measurements, as discussed in the next section.

\subsection{GB concentration: Simulations vs. experiments}
The estimated $X_\text{GB}$, accounting for the vibrational entropy contribution in simulated Mg alloy systems, were systematically compared across various temperature and bulk concentration regimes to experimentally measured local chemical distributions. These experimental measurements were obtained using APT or energy-dispersive X-ray spectroscopy (EDS).  The temperatures selected for comparison correspond to the thermomechanical temperatures prior to quenching and subsequent sample preparation for microscopic and chemical characterization in the experiments.

\begin{figure*}[hbt!]
\centering
\includegraphics[width=\textwidth]{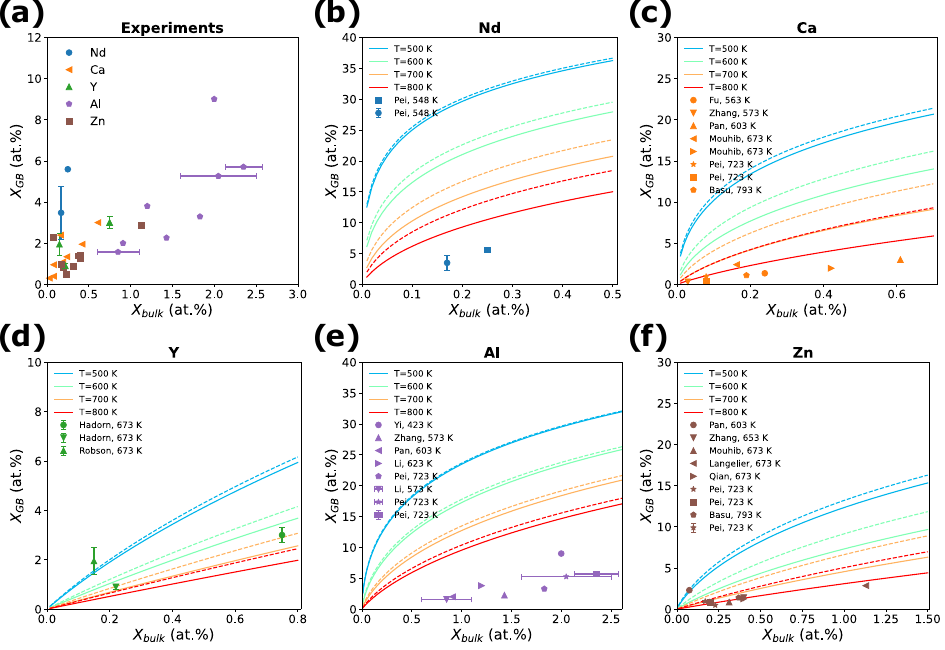}
\caption{(a) Summary of experimentally measured concentrations of various solutes at GB versus bulk in Mg alloys~\cite{pei2021grain,fu2022achieving,li2022elucidation,li2022elucidating,pei2022synergistic,yi2023interplay,zhang2023anisotropic,pan2020mechanistic,basu2022segregation,pei2022effect,hadorn2012role,robson2016grain,langelier2017effects,qian2022influence,zhang2022significantly,pei2023atomistic,mouhib2024exploring}. The concentrations of (b) Nd, (c) Ca, (d) Y, (e) Al, and (f) Zn solutes at GBs within the simulated Mg polycrystal as a function of their bulk solute concentrations, incorporating the vibrational entropy contribution $\Delta E_{\text{seg}}+\Delta F_{\text{seg}}^{\text{vib}}$ (represented by solid lines) and considering only the segregation energy $\Delta E_{\text{seg}}$ (represented by dashed lines) across various temperature regimes following the Langmuir–McLean model, as denoted in Equations~\eqref{equ11} and~\eqref{equ12}. Experimental data presented in (a) are illustrated as black dots on the corresponding graphs, labeled with the first authors' names and the thermomechanical processing temperatures of the Mg alloys.}
\label{fig7}
\end{figure*}

Experimental data of $X_\text{GB}$ versus $X_\text{bulk}$ are summarized in Figure~\ref{fig7}(a). Notably, the Nd solute exhibits significantly higher $X_\text{GB}$ than other solute elements under similar $X_\text{bulk}$ and temperature conditions. This experimental observation is in good agreement with the simulation results presented in Figure~\ref{fig3}, which demonstrate an elevated $X_\text{GB}$ for Nd at both 0.1 at.\% and 1 at.\% $X_\text{bulk}$ across the temperature range up to 800 K. This agreement between experimental and simulated data underscores Nd's exceptional energetic preference for segregation at GBs, establishing it as the most susceptible among the studied alloying elements to undergo segregation. However, it is important to note that the simulation predicts the $X_\text{GB}$ of Nd solute with an overestimation by a factor of 4 to 6 in the temperature regime from 500 to 600 K (see Figure~\ref{fig7}(b)). An overestimation in $X_\text{GB}$ for the Ca and Al solutes was also identified, with a factor of 2 to 3 across the relevant temperatures and $X_\text{bulk}$ regimes (see Figure~\ref{fig7}(c,e)).
A relatively good agreement was observed between experimental measurements and simulated predictions of $X_\text{GB}$ for Y and Zn solutes. The experimentally measured $X_\text{GB}$ values fall within the range predicted by atomistic simulations. To the best of our knowledge, no experimental segregation data on the chemical distribution of Li in GB were available. This is primarily attributed to the research focus on forming Mg-Li intermetallic phases to enhance the mechanical performances of Mg alloys, rather than on engineering GB properties below the solid solubility limit~\cite{chang2006mechanical,peng2022plastic}.

The overestimation of $X_\text{GB}$ could be attributed to factors related to both theoretical methodologies and experimental measurements, which are not regarded as inherent limitations of the approach. 
Firstly, the Langmuir–McLean model, employed in our current study, assumes an infinitesimally diluted solid solution, where solute-solute interactions are not accounted for. Experimental Mg alloy systems are often more complex than simple binary compositions, making it necessary to account for synergistic effects from solute-solute interactions. This simplification could potentially lead to disparities between experimental observations and theoretical predictions, e.g., as demonstrated in our recent APT study on the Mg-Zn-Ca alloy systems~\cite{mouhib2024exploring}. There, it was shown that the presence of Zn significantly reduced the concentration of Ca at GBs. This was evident from a decrease from 2.39 at.\% in a Mg-Ca binary system with $X_\text{bulk}$ of Ca at 0.163 at.\% to 1.95 at.\% in a Mg-Zn-Ca ternary system with even higher bulk Ca concentration of 0.421 at.\% (see Mouhib, 673 K in Figure~\ref{fig7}(c))~\cite{mouhib2024exploring}. Given that this binary alloy system for Ca segregation is the only example available in the literature, it is noteworthy that the experimentally measured $X_\text{GB}$ for this system demonstrates a closer alignment with theoretical predictions compared to results from ternary or quaternary alloy systems. Moreover, the available experimental data on Y segregation at GBs were all obtained from Mg-Y binary alloy systems (Figure~\ref{fig7}(d)), demonstrating a good correlation with theoretical predictions.
To investigate solute segregation behavior at higher concentrations, hybrid Monte-Carlo/molecular dynamics (MC/MD) approaches can be employed to provide thermodynamic insights into solute clustering, precipitation, and the formation of intermetallic phases at GBs~\cite{sadigh2012scalable,pan2016effect,ganguly2024grain}.

Additionally, the discrepancy may stem from the limitations in experimental measurements, which were performed on selected individual GBs, contrasting with atomistic simulations that encompassed a polycrystal with over 38,000 GB sites in a GB network with various macro and microscopic characteristics. As revealed in our recent APT study, significant variations were observed in the segregation behavior of Nd among GBs in Mg alloys, which possess distinct macroscopic characteristics~\cite{pei2023atomistic}. Notably, the concentration of Nd at GBs ($X_\text{GB}$) was found to be more than twice as high in a GB with a misorientation angle of 61.1$^{\circ}$ compared to one with an angle of 35.8$^{\circ}$. This scattering in GB segregation can primarily be attributed to variations in the local atomic environments, indicating the crucial role of microscopic characteristics in influencing solute distribution at the atomic level. Furthermore, in experimental conditions, there is always competition with other defects, such as dislocations or stacking faults, which cannot be switched off. If a large number of solutes get trapped in these defects, this can also lead to less solute being available to segregate at the GBs. This underscores the need for defect phase diagrams~\cite{korte2022defect} to better understand solute distribution mechanisms under different chemical potential conditions.

Furthermore, artifacts and approximations in both experimental and simulation methodologies can significantly impact the outcome solute segregation behavior. For instance, during the post-processing of APT and EDS data, assumptions about the GB thickness can lead to notable discrepancies in the measurements of $X_\text{GB}$. Similarly, the definition of GB sites according to structural analysis algorithms for segregation energy calculations influences the segregation energy spectrum and, consequently, the estimated solute concentration at GBs. In atomistic simulations, the choice of methodological assumptions for calculating $\Delta F_{\text{seg}}^{\text{vib}}$, such as those made by the HA method, can influence results. Specifically, the HA method’s inability to capture anharmonic behaviors—--like structural transitions—--and its omission of other significant entropy contributions, especially at elevated temperatures, are notable limitations of this model. The choice of interatomic potential also impacts the simulation outcomes. For instance, notable differences in segregation energy spectra were observed when comparing results obtained using the MEAM~\cite{kim2009atomistic} and EAM~\cite{mendelev2009development} potentials, see Figure S37. Nevertheless, both potentials yielded similar mean segregation energy values of approximately -0.08 eV/atom for Al, indicating a comparable overall tendency of Al solutes to segregate to GBs. The limitations of semi-empirical potentials in accurately capturing the variation in solute segregation behaviors across the diverse local atomic environments of GBs and a broad range of alloy systems have been reported~\cite{wagih2022learning}. Hybrid multiscale modeling approaches, such as quantum mechanical/molecular mechanical (QM/MM), could offer QM-level accurate results while circumventing the size constraints typically associated with ab-initio calculations.

\subsection{Generalizability of ML models}

In this study, we systematically investigated the segregation behavior of six solute elements commonly used in engineering Mg alloys within polycrystalline Mg. Unlike prior studies that primarily focused on individual highly symmetric GBs, such as symmetric tilt GBs and twin boundaries \cite{huber2014atomistic,messina2021machine,menon2024atomistic,wang2024defects}, this work encompasses a significantly broader sampling of local atomic environments. For instance, our previous work on basal-textured polycrystalline Mg \cite{ganguly2024grain}, which explored a range of \hkl<0001> symmetric tilt GBs, revealed that the segregation energy spectrum for such specific GBs exhibited distinct bimodal characteristics. 
By including general GBs and junctions, this study captures the diversity of sites present in realistic polycrystalline materials, thereby reflecting the heterogeneous nature of solute segregation. This expanded scope provides a more comprehensive understanding of thermodynamic segregation tendencies and enables stronger correlations with experimentally observed structure-property relationships in Mg alloys.

Moreover, previous ML models for solute segregation in Mg GBs were often limited to single elements, such as Al \cite{messina2021machine} or Y \cite{menon2024atomistic}, or used different interatomic potentials for the Mg matrix when handling multiple elements \cite{wagih2020learning}.  In contrast, this work considers the segregation behavior of multiple solutes simultaneously within a unified framework, which allows us to consider the effect of intrinsic properties of the solute elements on segregation tendencies. Segregation behavior at GBs is not only influenced by the characteristics of GB sites but also by the intrinsic properties of solutes, including atomic size, electronegativity, and bonding characteristics. Our findings highlight the variability in segregation energy spectra and feature importance rankings among the studied solutes, reflecting the complex interplay between solutes’ intrinsic properties and the local atomic environment at GB sites.

The ML models developed in this work utilize a combination of energetic and structural descriptors of the Mg matrix to predict segregation behavior. While these models were trained and validated using these six solute elements, their applicability to other elements in Mg alloys depends on the generalizability of the features used. The energetic and structural features, such as hydrostatic stress, atomic volume, and potential energy, serve as general descriptors that capture fundamental aspects of segregation behavior influenced by the characteristics of GB sites and are transferable across different solutes. Incorporating additional intrinsic features of solutes, such as misfit volume, cohesive energy, electronegativity differences, and diffusion coefficients, could further enhance the model’s ability to predict segregation behavior for a broader range of solutes by providing deeper insights into how solutes interact with the Mg matrix and local GB environments. 

Furthermore, expanding the training dataset to include a wider range of solute elements and more diverse local atomic environments is critical for improving the models’ robustness and generalizability. Performing cross-validation with experimental or computational data for additional solutes will ensure that the models capture broader trends and remain reliable for new predictions. These considerations underscore the potential for extending the current ML framework to predict segregation behavior across a broader spectrum of solute elements, ultimately providing a more comprehensive tool for understanding and engineering GB segregation in Mg alloys.

\FloatBarrier
\section{Conclusions}
In this study, the GB segregation behavior of Nd, Ca, Y, Li, Al, and Zn solute elements in polycrystalline Mg at 0 K and finite temperatures was systematically investigated using atomistic simulations. Machine learning (ML) models were employed to predict the segregation energy and free energy, incorporating energetic and structural descriptors. The GB solute concentrations in these Mg alloy systems were calculated following the Langmuir–McLean isotherm and compared with experimentally measured local chemical distributions at various temperature and bulk concentration regimes. Key insights from the study are summarized as follows: 

\begin{itemize}
  \item The distribution of vibrational free energy $F^{\text{vib}}$ approximates a normal distribution at finite temperatures in polycrystalline Mg. A notable deviation from the $F^{\text{vib}}$ value at bulk sites was observed, underscoring the substantial differences in thermal properties between GB and bulk regions.
  \item The spectra of segregation energy $\Delta E_{\text{seg}}$, segregation vibrational free energies $\Delta F_{\text{seg}}^{\text{vib}}$ and segregation free energy $\Delta F_{\text{seg}}$ at finite temperatures all exhibit skew-normal distributions in the simulated Mg alloys.
  \item ML models were successfully trained to predict $\Delta E_{\text{seg}}$ and $\Delta F_{\text{seg}}$ using energetic and structural descriptors, with hydrostatic stress demonstrating the highest importance. The performance in predicting $\Delta F_{\text{seg}}^{\text{vib}}$ was limited, indicating the need for incorporating more features that capture post-segregation dynamics.
  \item Flexibility volume ($V_{\text{flex}}$) provides a dimensionless measure of the local atomic environment's flexibility, contributing to the segregation tendency of solute atoms in addition to excess free volume.
  \item The Nd solute exhibits a significantly higher solute concentration $X_\text{GB}$ across temperature ranges up to 800 K compared to other solutes in this study, providing a rationale for the experimentally observed pronounced segregation tendency of Nd. 
  \item Among the solutes studied, the predicted $X_\text{GB}$ of Y and Zn closely match the experimental data. However, deviations observed in Nd, Al, and Ca at high bulk concentrations may be attributed to the limitations of the Langmuir–McLean model, which does not adequately consider solute-solute interactions, the definition of the GB region, and the accuracy of the interatomic potentials used.
  
\end{itemize}

\section*{Author contributions}
Z.X. performed atomistic simulations, applied machine learning models, and wrote the original draft. A.A. performed atomistic simulations and constructed local atomic environment descriptors. U.K. established the machine learning workflow. All authors contributed to the discussion and editing of the final manuscript.

\section*{Acknowledgements}

Z.X. and T.A.S. acknowledge the financial support by the German Research Foundation (DFG) (Grant Nr. 505716422). T.A.S. are grateful for the financial support from the DFG (Grant Nr. AL1343/7-1, AL1343/8-1 and Yi 103/3-1). Z.X., S.K.K. and U.K. acknowledge financial support by the DFG through the projects A05, A07 and C02 of the SFB1394 Structural and Chemical Atomic Complexity – From Defect Phase Diagrams to Material Properties, project ID 409476157. Additionally, Z.X. and S.K.K. are grateful for funding from the European Research Council (ERC) under the European Union’s Horizon 2020 research and innovation program (grant agreement No. 852096 FunBlocks). J.G. acknowledges funding from the French National Research Agency (ANR), Grant ANR-21-CE08-0001 (ATOUUM) and ANR-22-CE92-0058-01 (SILA). The authors gratefully acknowledge the computing time provided to them at the NHR Center NHR4CES at RWTH Aachen University (project number p0020431 and p0020267). This is funded by the Federal Ministry of Education and Research, and the state governments participating on the basis of the resolutions of the GWK for national high performance computing at universities (www.nhr-verein.de/unsere-partner). 

\section*{Competing interests}
All authors declare no financial or non-financial competing interests.

\section*{Data availability}
The datasets used and/or analyzed during the current study are available at https://doi.org/10.5281/zenodo.14605664.

\bibliography{main}

\begin{thebibliography}{86}%
\makeatletter
\providecommand \@ifxundefined [1]{%
 \@ifx{#1\undefined}
}%
\providecommand \@ifnum [1]{%
 \ifnum #1\expandafter \@firstoftwo
 \else \expandafter \@secondoftwo
 \fi
}%
\providecommand \@ifx [1]{%
 \ifx #1\expandafter \@firstoftwo
 \else \expandafter \@secondoftwo
 \fi
}%
\providecommand \natexlab [1]{#1}%
\providecommand \enquote  [1]{``#1''}%
\providecommand \bibnamefont  [1]{#1}%
\providecommand \bibfnamefont [1]{#1}%
\providecommand \citenamefont [1]{#1}%
\providecommand \href@noop [0]{\@secondoftwo}%
\providecommand \href [0]{\begingroup \@sanitize@url \@href}%
\providecommand \@href[1]{\@@startlink{#1}\@@href}%
\providecommand \@@href[1]{\endgroup#1\@@endlink}%
\providecommand \@sanitize@url [0]{\catcode `\\12\catcode `\$12\catcode `\&12\catcode `\#12\catcode `\^12\catcode `\_12\catcode `\%12\relax}%
\providecommand \@@startlink[1]{}%
\providecommand \@@endlink[0]{}%
\providecommand \url  [0]{\begingroup\@sanitize@url \@url }%
\providecommand \@url [1]{\endgroup\@href {#1}{\urlprefix }}%
\providecommand \urlprefix  [0]{URL }%
\providecommand \Eprint [0]{\href }%
\providecommand \doibase [0]{https://doi.org/}%
\providecommand \selectlanguage [0]{\@gobble}%
\providecommand \bibinfo  [0]{\@secondoftwo}%
\providecommand \bibfield  [0]{\@secondoftwo}%
\providecommand \translation [1]{[#1]}%
\providecommand \BibitemOpen [0]{}%
\providecommand \bibitemStop [0]{}%
\providecommand \bibitemNoStop [0]{.\EOS\space}%
\providecommand \EOS [0]{\spacefactor3000\relax}%
\providecommand \BibitemShut  [1]{\csname bibitem#1\endcsname}%
\let\auto@bib@innerbib\@empty
\bibitem [{\citenamefont {Mordike}\ and\ \citenamefont {Ebert}(2001)}]{mordike2001magnesium}%
  \BibitemOpen
  \bibfield  {author} {\bibinfo {author} {\bibfnamefont {B.}~\bibnamefont {Mordike}}\ and\ \bibinfo {author} {\bibfnamefont {T.}~\bibnamefont {Ebert}},\ }\bibfield  {title} {\bibinfo {title} {Magnesium: properties—applications—potential},\ }\href@noop {} {\bibfield  {journal} {\bibinfo  {journal} {Materials Science and Engineering: A}\ }\textbf {\bibinfo {volume} {302}},\ \bibinfo {pages} {37} (\bibinfo {year} {2001})}\BibitemShut {NoStop}%
\bibitem [{\citenamefont {Pollock}(2010)}]{pollock2010weight}%
  \BibitemOpen
  \bibfield  {author} {\bibinfo {author} {\bibfnamefont {T.~M.}\ \bibnamefont {Pollock}},\ }\bibfield  {title} {\bibinfo {title} {Weight loss with magnesium alloys},\ }\href@noop {} {\bibfield  {journal} {\bibinfo  {journal} {Science}\ }\textbf {\bibinfo {volume} {328}},\ \bibinfo {pages} {986} (\bibinfo {year} {2010})}\BibitemShut {NoStop}%
\bibitem [{\citenamefont {Yoo}(1981)}]{yoo1981slip}%
  \BibitemOpen
  \bibfield  {author} {\bibinfo {author} {\bibfnamefont {M.}~\bibnamefont {Yoo}},\ }\bibfield  {title} {\bibinfo {title} {Slip, twinning, and fracture in hexagonal close-packed metals},\ }\href@noop {} {\bibfield  {journal} {\bibinfo  {journal} {Metallurgical transactions A}\ }\textbf {\bibinfo {volume} {12}},\ \bibinfo {pages} {409} (\bibinfo {year} {1981})}\BibitemShut {NoStop}%
\bibitem [{\citenamefont {Kocks}\ \emph {et~al.}(2000)\citenamefont {Kocks}, \citenamefont {Tom{\'e}},\ and\ \citenamefont {Wenk}}]{kocks2000texture}%
  \BibitemOpen
  \bibfield  {author} {\bibinfo {author} {\bibfnamefont {U.~F.}\ \bibnamefont {Kocks}}, \bibinfo {author} {\bibfnamefont {C.~N.}\ \bibnamefont {Tom{\'e}}},\ and\ \bibinfo {author} {\bibfnamefont {H.-R.}\ \bibnamefont {Wenk}},\ }\href@noop {} {\emph {\bibinfo {title} {Texture and anisotropy: preferred orientations in polycrystals and their effect on materials properties}}}\ (\bibinfo  {publisher} {Cambridge university press},\ \bibinfo {year} {2000})\BibitemShut {NoStop}%
\bibitem [{\citenamefont {Song}\ \emph {et~al.}(2020)\citenamefont {Song}, \citenamefont {She}, \citenamefont {Chen},\ and\ \citenamefont {Pan}}]{song2020latest}%
  \BibitemOpen
  \bibfield  {author} {\bibinfo {author} {\bibfnamefont {J.}~\bibnamefont {Song}}, \bibinfo {author} {\bibfnamefont {J.}~\bibnamefont {She}}, \bibinfo {author} {\bibfnamefont {D.}~\bibnamefont {Chen}},\ and\ \bibinfo {author} {\bibfnamefont {F.}~\bibnamefont {Pan}},\ }\bibfield  {title} {\bibinfo {title} {Latest research advances on magnesium and magnesium alloys worldwide},\ }\href@noop {} {\bibfield  {journal} {\bibinfo  {journal} {Journal of Magnesium and Alloys}\ }\textbf {\bibinfo {volume} {8}},\ \bibinfo {pages} {1} (\bibinfo {year} {2020})}\BibitemShut {NoStop}%
\bibitem [{\citenamefont {Prasad}\ \emph {et~al.}(2022)\citenamefont {Prasad}, \citenamefont {Prasad}, \citenamefont {Verma}, \citenamefont {Mishra}, \citenamefont {Kumar},\ and\ \citenamefont {Singh}}]{prasad2022role}%
  \BibitemOpen
  \bibfield  {author} {\bibinfo {author} {\bibfnamefont {S.~S.}\ \bibnamefont {Prasad}}, \bibinfo {author} {\bibfnamefont {S.}~\bibnamefont {Prasad}}, \bibinfo {author} {\bibfnamefont {K.}~\bibnamefont {Verma}}, \bibinfo {author} {\bibfnamefont {R.~K.}\ \bibnamefont {Mishra}}, \bibinfo {author} {\bibfnamefont {V.}~\bibnamefont {Kumar}},\ and\ \bibinfo {author} {\bibfnamefont {S.}~\bibnamefont {Singh}},\ }\bibfield  {title} {\bibinfo {title} {The role and significance of magnesium in modern day research-a review},\ }\href@noop {} {\bibfield  {journal} {\bibinfo  {journal} {Journal of Magnesium and alloys}\ }\textbf {\bibinfo {volume} {10}},\ \bibinfo {pages} {1} (\bibinfo {year} {2022})}\BibitemShut {NoStop}%
\bibitem [{\citenamefont {Bai}\ \emph {et~al.}(2023)\citenamefont {Bai}, \citenamefont {Yang}, \citenamefont {Wen}, \citenamefont {Chen}, \citenamefont {Zhou}, \citenamefont {Jiang}, \citenamefont {Peng},\ and\ \citenamefont {Pan}}]{bai2023applications}%
  \BibitemOpen
  \bibfield  {author} {\bibinfo {author} {\bibfnamefont {J.}~\bibnamefont {Bai}}, \bibinfo {author} {\bibfnamefont {Y.}~\bibnamefont {Yang}}, \bibinfo {author} {\bibfnamefont {C.}~\bibnamefont {Wen}}, \bibinfo {author} {\bibfnamefont {J.}~\bibnamefont {Chen}}, \bibinfo {author} {\bibfnamefont {G.}~\bibnamefont {Zhou}}, \bibinfo {author} {\bibfnamefont {B.}~\bibnamefont {Jiang}}, \bibinfo {author} {\bibfnamefont {X.}~\bibnamefont {Peng}},\ and\ \bibinfo {author} {\bibfnamefont {F.}~\bibnamefont {Pan}},\ }\bibfield  {title} {\bibinfo {title} {Applications of magnesium alloys for aerospace: A review},\ }\href@noop {} {\bibfield  {journal} {\bibinfo  {journal} {Journal of Magnesium and Alloys}\ }\textbf {\bibinfo {volume} {11}},\ \bibinfo {pages} {3609} (\bibinfo {year} {2023})}\BibitemShut {NoStop}%
\bibitem [{\citenamefont {Sandl{\"o}bes}\ \emph {et~al.}(2011)\citenamefont {Sandl{\"o}bes}, \citenamefont {Zaefferer}, \citenamefont {Schestakow}, \citenamefont {Yi},\ and\ \citenamefont {Gonzalez-Martinez}}]{sandlobes2011role}%
  \BibitemOpen
  \bibfield  {author} {\bibinfo {author} {\bibfnamefont {S.}~\bibnamefont {Sandl{\"o}bes}}, \bibinfo {author} {\bibfnamefont {S.}~\bibnamefont {Zaefferer}}, \bibinfo {author} {\bibfnamefont {I.}~\bibnamefont {Schestakow}}, \bibinfo {author} {\bibfnamefont {S.}~\bibnamefont {Yi}},\ and\ \bibinfo {author} {\bibfnamefont {R.}~\bibnamefont {Gonzalez-Martinez}},\ }\bibfield  {title} {\bibinfo {title} {On the role of non-basal deformation mechanisms for the ductility of {M}g and {M}g--{Y} alloys},\ }\href@noop {} {\bibfield  {journal} {\bibinfo  {journal} {Acta Materialia}\ }\textbf {\bibinfo {volume} {59}},\ \bibinfo {pages} {429} (\bibinfo {year} {2011})}\BibitemShut {NoStop}%
\bibitem [{\citenamefont {Wang}\ and\ \citenamefont {Liu}(2022)}]{wang2022optimizing}%
  \BibitemOpen
  \bibfield  {author} {\bibinfo {author} {\bibfnamefont {T.}~\bibnamefont {Wang}}\ and\ \bibinfo {author} {\bibfnamefont {F.}~\bibnamefont {Liu}},\ }\bibfield  {title} {\bibinfo {title} {Optimizing mechanical properties of magnesium alloys by philosophy of thermo-kinetic synergy: Review and outlook},\ }\href@noop {} {\bibfield  {journal} {\bibinfo  {journal} {Journal of Magnesium and Alloys}\ }\textbf {\bibinfo {volume} {10}},\ \bibinfo {pages} {326} (\bibinfo {year} {2022})}\BibitemShut {NoStop}%
\bibitem [{\citenamefont {Ovri}\ \emph {et~al.}(2023)\citenamefont {Ovri}, \citenamefont {Markmann}, \citenamefont {Barthel}, \citenamefont {Kruth}, \citenamefont {Dieringa},\ and\ \citenamefont {Lilleodden}}]{ovri2023mechanistic}%
  \BibitemOpen
  \bibfield  {author} {\bibinfo {author} {\bibfnamefont {H.}~\bibnamefont {Ovri}}, \bibinfo {author} {\bibfnamefont {J.}~\bibnamefont {Markmann}}, \bibinfo {author} {\bibfnamefont {J.}~\bibnamefont {Barthel}}, \bibinfo {author} {\bibfnamefont {M.}~\bibnamefont {Kruth}}, \bibinfo {author} {\bibfnamefont {H.}~\bibnamefont {Dieringa}},\ and\ \bibinfo {author} {\bibfnamefont {E.~T.}\ \bibnamefont {Lilleodden}},\ }\bibfield  {title} {\bibinfo {title} {Mechanistic origin of the enhanced strength and ductility in mg-rare earth alloys},\ }\href@noop {} {\bibfield  {journal} {\bibinfo  {journal} {Acta Materialia}\ }\textbf {\bibinfo {volume} {244}},\ \bibinfo {pages} {118550} (\bibinfo {year} {2023})}\BibitemShut {NoStop}%
\bibitem [{\citenamefont {Basu}\ and\ \citenamefont {Al-Samman}(2014)}]{basu2014triggering}%
  \BibitemOpen
  \bibfield  {author} {\bibinfo {author} {\bibfnamefont {I.}~\bibnamefont {Basu}}\ and\ \bibinfo {author} {\bibfnamefont {T.}~\bibnamefont {Al-Samman}},\ }\bibfield  {title} {\bibinfo {title} {Triggering rare earth texture modification in magnesium alloys by addition of zinc and zirconium},\ }\href@noop {} {\bibfield  {journal} {\bibinfo  {journal} {Acta Materialia}\ }\textbf {\bibinfo {volume} {67}},\ \bibinfo {pages} {116} (\bibinfo {year} {2014})}\BibitemShut {NoStop}%
\bibitem [{\citenamefont {Trang}\ \emph {et~al.}(2018)\citenamefont {Trang}, \citenamefont {Zhang}, \citenamefont {Kim}, \citenamefont {Zargaran}, \citenamefont {Hwang}, \citenamefont {Suh},\ and\ \citenamefont {Kim}}]{trang2018designing}%
  \BibitemOpen
  \bibfield  {author} {\bibinfo {author} {\bibfnamefont {T.}~\bibnamefont {Trang}}, \bibinfo {author} {\bibfnamefont {J.}~\bibnamefont {Zhang}}, \bibinfo {author} {\bibfnamefont {J.}~\bibnamefont {Kim}}, \bibinfo {author} {\bibfnamefont {A.}~\bibnamefont {Zargaran}}, \bibinfo {author} {\bibfnamefont {J.}~\bibnamefont {Hwang}}, \bibinfo {author} {\bibfnamefont {B.-C.}\ \bibnamefont {Suh}},\ and\ \bibinfo {author} {\bibfnamefont {N.}~\bibnamefont {Kim}},\ }\bibfield  {title} {\bibinfo {title} {Designing a magnesium alloy with high strength and high formability},\ }\href@noop {} {\bibfield  {journal} {\bibinfo  {journal} {Nature communications}\ }\textbf {\bibinfo {volume} {9}},\ \bibinfo {pages} {2522} (\bibinfo {year} {2018})}\BibitemShut {NoStop}%
\bibitem [{\citenamefont {Chen}\ \emph {et~al.}(2022)\citenamefont {Chen}, \citenamefont {Chen}, \citenamefont {Su}, \citenamefont {He}, \citenamefont {Tan}, \citenamefont {Xu}, \citenamefont {Huang}, \citenamefont {Dai},\ and\ \citenamefont {Lu}}]{chen2022mechanisms}%
  \BibitemOpen
  \bibfield  {author} {\bibinfo {author} {\bibfnamefont {Q.}~\bibnamefont {Chen}}, \bibinfo {author} {\bibfnamefont {R.}~\bibnamefont {Chen}}, \bibinfo {author} {\bibfnamefont {J.}~\bibnamefont {Su}}, \bibinfo {author} {\bibfnamefont {Q.}~\bibnamefont {He}}, \bibinfo {author} {\bibfnamefont {B.}~\bibnamefont {Tan}}, \bibinfo {author} {\bibfnamefont {C.}~\bibnamefont {Xu}}, \bibinfo {author} {\bibfnamefont {X.}~\bibnamefont {Huang}}, \bibinfo {author} {\bibfnamefont {Q.}~\bibnamefont {Dai}},\ and\ \bibinfo {author} {\bibfnamefont {J.}~\bibnamefont {Lu}},\ }\bibfield  {title} {\bibinfo {title} {The mechanisms of grain growth of {M}g alloys: a review},\ }\href@noop {} {\bibfield  {journal} {\bibinfo  {journal} {Journal of Magnesium and Alloys}\ }\textbf {\bibinfo {volume} {10}},\ \bibinfo {pages} {2384} (\bibinfo {year} {2022})}\BibitemShut {NoStop}%
\bibitem [{\citenamefont {Zeng}\ \emph {et~al.}(2016)\citenamefont {Zeng}, \citenamefont {Zhu}, \citenamefont {Xu}, \citenamefont {Bian}, \citenamefont {Davies}, \citenamefont {Birbilis},\ and\ \citenamefont {Nie}}]{zeng2016texture}%
  \BibitemOpen
  \bibfield  {author} {\bibinfo {author} {\bibfnamefont {Z.}~\bibnamefont {Zeng}}, \bibinfo {author} {\bibfnamefont {Y.}~\bibnamefont {Zhu}}, \bibinfo {author} {\bibfnamefont {S.}~\bibnamefont {Xu}}, \bibinfo {author} {\bibfnamefont {M.}~\bibnamefont {Bian}}, \bibinfo {author} {\bibfnamefont {C.}~\bibnamefont {Davies}}, \bibinfo {author} {\bibfnamefont {N.}~\bibnamefont {Birbilis}},\ and\ \bibinfo {author} {\bibfnamefont {J.}~\bibnamefont {Nie}},\ }\bibfield  {title} {\bibinfo {title} {Texture evolution during static recrystallization of cold-rolled magnesium alloys},\ }\href@noop {} {\bibfield  {journal} {\bibinfo  {journal} {Acta Materialia}\ }\textbf {\bibinfo {volume} {105}},\ \bibinfo {pages} {479} (\bibinfo {year} {2016})}\BibitemShut {NoStop}%
\bibitem [{\citenamefont {Barrett}\ \emph {et~al.}(2017)\citenamefont {Barrett}, \citenamefont {Imandoust}, \citenamefont {Oppedal}, \citenamefont {Inal}, \citenamefont {Tschopp},\ and\ \citenamefont {El~Kadiri}}]{barrett2017effect}%
  \BibitemOpen
  \bibfield  {author} {\bibinfo {author} {\bibfnamefont {C.~D.}\ \bibnamefont {Barrett}}, \bibinfo {author} {\bibfnamefont {A.}~\bibnamefont {Imandoust}}, \bibinfo {author} {\bibfnamefont {A.~L.}\ \bibnamefont {Oppedal}}, \bibinfo {author} {\bibfnamefont {K.}~\bibnamefont {Inal}}, \bibinfo {author} {\bibfnamefont {M.~A.}\ \bibnamefont {Tschopp}},\ and\ \bibinfo {author} {\bibfnamefont {H.}~\bibnamefont {El~Kadiri}},\ }\bibfield  {title} {\bibinfo {title} {Effect of grain boundaries on texture formation during dynamic recrystallization of magnesium alloys},\ }\href@noop {} {\bibfield  {journal} {\bibinfo  {journal} {Acta Materialia}\ }\textbf {\bibinfo {volume} {128}},\ \bibinfo {pages} {270} (\bibinfo {year} {2017})}\BibitemShut {NoStop}%
\bibitem [{\citenamefont {Nie}\ \emph {et~al.}(2013)\citenamefont {Nie}, \citenamefont {Zhu}, \citenamefont {Liu},\ and\ \citenamefont {Fang}}]{nie2013periodic}%
  \BibitemOpen
  \bibfield  {author} {\bibinfo {author} {\bibfnamefont {J.~F.}\ \bibnamefont {Nie}}, \bibinfo {author} {\bibfnamefont {Y.}~\bibnamefont {Zhu}}, \bibinfo {author} {\bibfnamefont {J.}~\bibnamefont {Liu}},\ and\ \bibinfo {author} {\bibfnamefont {X.-Y.}\ \bibnamefont {Fang}},\ }\bibfield  {title} {\bibinfo {title} {Periodic segregation of solute atoms in fully coherent twin boundaries},\ }\href@noop {} {\bibfield  {journal} {\bibinfo  {journal} {Science}\ }\textbf {\bibinfo {volume} {340}},\ \bibinfo {pages} {957} (\bibinfo {year} {2013})}\BibitemShut {NoStop}%
\bibitem [{\citenamefont {Zhou}\ \emph {et~al.}(2015)\citenamefont {Zhou}, \citenamefont {Cheng}, \citenamefont {Ma}, \citenamefont {Xu}, \citenamefont {Mathaudhu}, \citenamefont {Wang},\ and\ \citenamefont {Zhu}}]{zhou2015effect}%
  \BibitemOpen
  \bibfield  {author} {\bibinfo {author} {\bibfnamefont {H.}~\bibnamefont {Zhou}}, \bibinfo {author} {\bibfnamefont {G.}~\bibnamefont {Cheng}}, \bibinfo {author} {\bibfnamefont {X.}~\bibnamefont {Ma}}, \bibinfo {author} {\bibfnamefont {W.}~\bibnamefont {Xu}}, \bibinfo {author} {\bibfnamefont {S.}~\bibnamefont {Mathaudhu}}, \bibinfo {author} {\bibfnamefont {Q.}~\bibnamefont {Wang}},\ and\ \bibinfo {author} {\bibfnamefont {Y.}~\bibnamefont {Zhu}},\ }\bibfield  {title} {\bibinfo {title} {Effect of {A}g on interfacial segregation in {M}g--{G}d--{Y}--({A}g)--{Z}r alloy},\ }\href@noop {} {\bibfield  {journal} {\bibinfo  {journal} {Acta Materialia}\ }\textbf {\bibinfo {volume} {95}},\ \bibinfo {pages} {20} (\bibinfo {year} {2015})}\BibitemShut {NoStop}%
\bibitem [{\citenamefont {He}\ \emph {et~al.}(2021)\citenamefont {He}, \citenamefont {Li}, \citenamefont {Chen}, \citenamefont {Wilson},\ and\ \citenamefont {Nie}}]{he2021unusual}%
  \BibitemOpen
  \bibfield  {author} {\bibinfo {author} {\bibfnamefont {C.}~\bibnamefont {He}}, \bibinfo {author} {\bibfnamefont {Z.}~\bibnamefont {Li}}, \bibinfo {author} {\bibfnamefont {H.}~\bibnamefont {Chen}}, \bibinfo {author} {\bibfnamefont {N.}~\bibnamefont {Wilson}},\ and\ \bibinfo {author} {\bibfnamefont {J.-F.}\ \bibnamefont {Nie}},\ }\bibfield  {title} {\bibinfo {title} {Unusual solute segregation phenomenon in coherent twin boundaries},\ }\href@noop {} {\bibfield  {journal} {\bibinfo  {journal} {Nature communications}\ }\textbf {\bibinfo {volume} {12}},\ \bibinfo {pages} {722} (\bibinfo {year} {2021})}\BibitemShut {NoStop}%
\bibitem [{\citenamefont {Xie}\ \emph {et~al.}(2021)\citenamefont {Xie}, \citenamefont {Huang}, \citenamefont {Bai}, \citenamefont {Li}, \citenamefont {Liu}, \citenamefont {Feng}, \citenamefont {Yang}, \citenamefont {Pan}, \citenamefont {Li}, \citenamefont {Ren} \emph {et~al.}}]{xie2021nonsymmetrical}%
  \BibitemOpen
  \bibfield  {author} {\bibinfo {author} {\bibfnamefont {H.}~\bibnamefont {Xie}}, \bibinfo {author} {\bibfnamefont {Q.}~\bibnamefont {Huang}}, \bibinfo {author} {\bibfnamefont {J.}~\bibnamefont {Bai}}, \bibinfo {author} {\bibfnamefont {S.}~\bibnamefont {Li}}, \bibinfo {author} {\bibfnamefont {Y.}~\bibnamefont {Liu}}, \bibinfo {author} {\bibfnamefont {J.}~\bibnamefont {Feng}}, \bibinfo {author} {\bibfnamefont {Y.}~\bibnamefont {Yang}}, \bibinfo {author} {\bibfnamefont {H.}~\bibnamefont {Pan}}, \bibinfo {author} {\bibfnamefont {H.}~\bibnamefont {Li}}, \bibinfo {author} {\bibfnamefont {Y.}~\bibnamefont {Ren}}, \emph {et~al.},\ }\bibfield  {title} {\bibinfo {title} {Nonsymmetrical segregation of solutes in periodic misfit dislocations separated tilt grain boundaries},\ }\href@noop {} {\bibfield  {journal} {\bibinfo  {journal} {Nano Letters}\ }\textbf {\bibinfo {volume} {21}},\ \bibinfo {pages} {2870} (\bibinfo {year} {2021})}\BibitemShut {NoStop}%
\bibitem [{\citenamefont {Pei}\ \emph {et~al.}(2021)\citenamefont {Pei}, \citenamefont {Zou}, \citenamefont {Wei},\ and\ \citenamefont {Al-Samman}}]{pei2021grain}%
  \BibitemOpen
  \bibfield  {author} {\bibinfo {author} {\bibfnamefont {R.}~\bibnamefont {Pei}}, \bibinfo {author} {\bibfnamefont {Y.}~\bibnamefont {Zou}}, \bibinfo {author} {\bibfnamefont {D.}~\bibnamefont {Wei}},\ and\ \bibinfo {author} {\bibfnamefont {T.}~\bibnamefont {Al-Samman}},\ }\bibfield  {title} {\bibinfo {title} {Grain boundary co-segregation in magnesium alloys with multiple substitutional elements},\ }\href@noop {} {\bibfield  {journal} {\bibinfo  {journal} {Acta Materialia}\ }\textbf {\bibinfo {volume} {208}},\ \bibinfo {pages} {116749} (\bibinfo {year} {2021})}\BibitemShut {NoStop}%
\bibitem [{\citenamefont {Fu}\ \emph {et~al.}(2022)\citenamefont {Fu}, \citenamefont {Sun}, \citenamefont {Ge}, \citenamefont {Xie}, \citenamefont {Li}, \citenamefont {Pan},\ and\ \citenamefont {Qin}}]{fu2022achieving}%
  \BibitemOpen
  \bibfield  {author} {\bibinfo {author} {\bibfnamefont {T.}~\bibnamefont {Fu}}, \bibinfo {author} {\bibfnamefont {X.}~\bibnamefont {Sun}}, \bibinfo {author} {\bibfnamefont {C.}~\bibnamefont {Ge}}, \bibinfo {author} {\bibfnamefont {D.}~\bibnamefont {Xie}}, \bibinfo {author} {\bibfnamefont {J.}~\bibnamefont {Li}}, \bibinfo {author} {\bibfnamefont {H.}~\bibnamefont {Pan}},\ and\ \bibinfo {author} {\bibfnamefont {G.}~\bibnamefont {Qin}},\ }\bibfield  {title} {\bibinfo {title} {Achieving high strength-ductility synergy in dilute {M}g-{A}l-{C}a alloy by trace ce addition},\ }\href@noop {} {\bibfield  {journal} {\bibinfo  {journal} {Journal of Alloys and Compounds}\ }\textbf {\bibinfo {volume} {917}},\ \bibinfo {pages} {165407} (\bibinfo {year} {2022})}\BibitemShut {NoStop}%
\bibitem [{\citenamefont {Li}\ \emph {et~al.}(2022{\natexlab{a}})\citenamefont {Li}, \citenamefont {Zhou}, \citenamefont {Breen}, \citenamefont {Peng}, \citenamefont {Su}, \citenamefont {K{\"u}rnsteiner}, \citenamefont {Correa}, \citenamefont {Chwa{\l}ek}, \citenamefont {Wang}, \citenamefont {Holec} \emph {et~al.}}]{li2022elucidation}%
  \BibitemOpen
  \bibfield  {author} {\bibinfo {author} {\bibfnamefont {J.}~\bibnamefont {Li}}, \bibinfo {author} {\bibfnamefont {X.}~\bibnamefont {Zhou}}, \bibinfo {author} {\bibfnamefont {A.}~\bibnamefont {Breen}}, \bibinfo {author} {\bibfnamefont {Z.}~\bibnamefont {Peng}}, \bibinfo {author} {\bibfnamefont {J.}~\bibnamefont {Su}}, \bibinfo {author} {\bibfnamefont {P.}~\bibnamefont {K{\"u}rnsteiner}}, \bibinfo {author} {\bibfnamefont {M.~J.~D.}\ \bibnamefont {Correa}}, \bibinfo {author} {\bibfnamefont {M.~L.}\ \bibnamefont {Chwa{\l}ek}}, \bibinfo {author} {\bibfnamefont {H.}~\bibnamefont {Wang}}, \bibinfo {author} {\bibfnamefont {D.}~\bibnamefont {Holec}}, \emph {et~al.},\ }\bibfield  {title} {\bibinfo {title} {Elucidation of formation and transformation mechanisms of {C}a-rich laves phase in {M}g-{A}l-{C}a-{M}n alloys},\ }\href@noop {} {\bibfield  {journal} {\bibinfo  {journal} {Journal of alloys and compounds}\ }\textbf {\bibinfo {volume} {928}},\ \bibinfo {pages} {167177} (\bibinfo {year}
  {2022}{\natexlab{a}})}\BibitemShut {NoStop}%
\bibitem [{\citenamefont {Li}\ \emph {et~al.}(2022{\natexlab{b}})\citenamefont {Li}, \citenamefont {Zhou}, \citenamefont {Su}, \citenamefont {Breitbach}, \citenamefont {Chwa{\l}ek}, \citenamefont {Wang},\ and\ \citenamefont {Dehm}}]{li2022elucidating}%
  \BibitemOpen
  \bibfield  {author} {\bibinfo {author} {\bibfnamefont {J.}~\bibnamefont {Li}}, \bibinfo {author} {\bibfnamefont {X.}~\bibnamefont {Zhou}}, \bibinfo {author} {\bibfnamefont {J.}~\bibnamefont {Su}}, \bibinfo {author} {\bibfnamefont {B.}~\bibnamefont {Breitbach}}, \bibinfo {author} {\bibfnamefont {M.~L.}\ \bibnamefont {Chwa{\l}ek}}, \bibinfo {author} {\bibfnamefont {H.}~\bibnamefont {Wang}},\ and\ \bibinfo {author} {\bibfnamefont {G.}~\bibnamefont {Dehm}},\ }\bibfield  {title} {\bibinfo {title} {Elucidating dynamic precipitation and yield strength of rolled {M}g--{A}l--{C}a--{M}n alloy},\ }\href@noop {} {\bibfield  {journal} {\bibinfo  {journal} {Materials Science and Engineering: A}\ }\textbf {\bibinfo {volume} {856}},\ \bibinfo {pages} {143898} (\bibinfo {year} {2022}{\natexlab{b}})}\BibitemShut {NoStop}%
\bibitem [{\citenamefont {Pei}\ \emph {et~al.}(2022{\natexlab{a}})\citenamefont {Pei}, \citenamefont {Zou}, \citenamefont {Zubair}, \citenamefont {Wei},\ and\ \citenamefont {Al-Samman}}]{pei2022synergistic}%
  \BibitemOpen
  \bibfield  {author} {\bibinfo {author} {\bibfnamefont {R.}~\bibnamefont {Pei}}, \bibinfo {author} {\bibfnamefont {Y.}~\bibnamefont {Zou}}, \bibinfo {author} {\bibfnamefont {M.}~\bibnamefont {Zubair}}, \bibinfo {author} {\bibfnamefont {D.}~\bibnamefont {Wei}},\ and\ \bibinfo {author} {\bibfnamefont {T.}~\bibnamefont {Al-Samman}},\ }\bibfield  {title} {\bibinfo {title} {Synergistic effect of {Y} and {C}a addition on the texture modification in {AZ31B} magnesium alloy},\ }\href@noop {} {\bibfield  {journal} {\bibinfo  {journal} {Acta Materialia}\ }\textbf {\bibinfo {volume} {233}},\ \bibinfo {pages} {117990} (\bibinfo {year} {2022}{\natexlab{a}})}\BibitemShut {NoStop}%
\bibitem [{\citenamefont {Yi}\ \emph {et~al.}(2023)\citenamefont {Yi}, \citenamefont {Sasaki}, \citenamefont {Prameela}, \citenamefont {Weihs},\ and\ \citenamefont {Falk}}]{yi2023interplay}%
  \BibitemOpen
  \bibfield  {author} {\bibinfo {author} {\bibfnamefont {P.}~\bibnamefont {Yi}}, \bibinfo {author} {\bibfnamefont {T.~T.}\ \bibnamefont {Sasaki}}, \bibinfo {author} {\bibfnamefont {S.~E.}\ \bibnamefont {Prameela}}, \bibinfo {author} {\bibfnamefont {T.~P.}\ \bibnamefont {Weihs}},\ and\ \bibinfo {author} {\bibfnamefont {M.~L.}\ \bibnamefont {Falk}},\ }\bibfield  {title} {\bibinfo {title} {The interplay between solute atoms and vacancy clusters in magnesium alloys},\ }\href@noop {} {\bibfield  {journal} {\bibinfo  {journal} {Acta Materialia}\ }\textbf {\bibinfo {volume} {249}},\ \bibinfo {pages} {118805} (\bibinfo {year} {2023})}\BibitemShut {NoStop}%
\bibitem [{\citenamefont {Zhang}\ \emph {et~al.}(2023)\citenamefont {Zhang}, \citenamefont {Jia}, \citenamefont {Zha}, \citenamefont {Zhao}, \citenamefont {Hua}, \citenamefont {Wang}, \citenamefont {Gao},\ and\ \citenamefont {Wang}}]{zhang2023anisotropic}%
  \BibitemOpen
  \bibfield  {author} {\bibinfo {author} {\bibfnamefont {M.-N.}\ \bibnamefont {Zhang}}, \bibinfo {author} {\bibfnamefont {H.-L.}\ \bibnamefont {Jia}}, \bibinfo {author} {\bibfnamefont {M.}~\bibnamefont {Zha}}, \bibinfo {author} {\bibfnamefont {L.}~\bibnamefont {Zhao}}, \bibinfo {author} {\bibfnamefont {Z.-M.}\ \bibnamefont {Hua}}, \bibinfo {author} {\bibfnamefont {C.}~\bibnamefont {Wang}}, \bibinfo {author} {\bibfnamefont {Y.-P.}\ \bibnamefont {Gao}},\ and\ \bibinfo {author} {\bibfnamefont {H.-Y.}\ \bibnamefont {Wang}},\ }\bibfield  {title} {\bibinfo {title} {Anisotropic segregation-driven texture weakening in a dilute {M}g-{A}l-{C}a alloy during isothermal annealing},\ }\href@noop {} {\bibfield  {journal} {\bibinfo  {journal} {Materials Research Letters}\ }\textbf {\bibinfo {volume} {11}},\ \bibinfo {pages} {781} (\bibinfo {year} {2023})}\BibitemShut {NoStop}%
\bibitem [{\citenamefont {Pan}\ \emph {et~al.}(2020)\citenamefont {Pan}, \citenamefont {Kang}, \citenamefont {Li}, \citenamefont {Xie}, \citenamefont {Zeng}, \citenamefont {Huang}, \citenamefont {Yang}, \citenamefont {Ren},\ and\ \citenamefont {Qin}}]{pan2020mechanistic}%
  \BibitemOpen
  \bibfield  {author} {\bibinfo {author} {\bibfnamefont {H.}~\bibnamefont {Pan}}, \bibinfo {author} {\bibfnamefont {R.}~\bibnamefont {Kang}}, \bibinfo {author} {\bibfnamefont {J.}~\bibnamefont {Li}}, \bibinfo {author} {\bibfnamefont {H.}~\bibnamefont {Xie}}, \bibinfo {author} {\bibfnamefont {Z.}~\bibnamefont {Zeng}}, \bibinfo {author} {\bibfnamefont {Q.}~\bibnamefont {Huang}}, \bibinfo {author} {\bibfnamefont {C.}~\bibnamefont {Yang}}, \bibinfo {author} {\bibfnamefont {Y.}~\bibnamefont {Ren}},\ and\ \bibinfo {author} {\bibfnamefont {G.}~\bibnamefont {Qin}},\ }\bibfield  {title} {\bibinfo {title} {Mechanistic investigation of a low-alloy {M}g--{C}a-based extrusion alloy with high strength--ductility synergy},\ }\href@noop {} {\bibfield  {journal} {\bibinfo  {journal} {Acta Materialia}\ }\textbf {\bibinfo {volume} {186}},\ \bibinfo {pages} {278} (\bibinfo {year} {2020})}\BibitemShut {NoStop}%
\bibitem [{\citenamefont {Basu}\ \emph {et~al.}(2022)\citenamefont {Basu}, \citenamefont {Chen}, \citenamefont {Wheeler}, \citenamefont {Sch{\"a}ublin},\ and\ \citenamefont {L{\"o}ffler}}]{basu2022segregation}%
  \BibitemOpen
  \bibfield  {author} {\bibinfo {author} {\bibfnamefont {I.}~\bibnamefont {Basu}}, \bibinfo {author} {\bibfnamefont {M.}~\bibnamefont {Chen}}, \bibinfo {author} {\bibfnamefont {J.}~\bibnamefont {Wheeler}}, \bibinfo {author} {\bibfnamefont {R.}~\bibnamefont {Sch{\"a}ublin}},\ and\ \bibinfo {author} {\bibfnamefont {J.~F.}\ \bibnamefont {L{\"o}ffler}},\ }\bibfield  {title} {\bibinfo {title} {Segregation-driven exceptional twin-boundary strengthening in lean {M}g--{Z}n--{C}a alloys},\ }\href@noop {} {\bibfield  {journal} {\bibinfo  {journal} {Acta Materialia}\ }\textbf {\bibinfo {volume} {229}},\ \bibinfo {pages} {117746} (\bibinfo {year} {2022})}\BibitemShut {NoStop}%
\bibitem [{\citenamefont {Pei}\ \emph {et~al.}(2022{\natexlab{b}})\citenamefont {Pei}, \citenamefont {Woo}, \citenamefont {Yi},\ and\ \citenamefont {Al-Samman}}]{pei2022effect}%
  \BibitemOpen
  \bibfield  {author} {\bibinfo {author} {\bibfnamefont {R.}~\bibnamefont {Pei}}, \bibinfo {author} {\bibfnamefont {S.~K.}\ \bibnamefont {Woo}}, \bibinfo {author} {\bibfnamefont {S.}~\bibnamefont {Yi}},\ and\ \bibinfo {author} {\bibfnamefont {T.}~\bibnamefont {Al-Samman}},\ }\bibfield  {title} {\bibinfo {title} {Effect of solute clusters on plastic instability in magnesium alloys},\ }\href@noop {} {\bibfield  {journal} {\bibinfo  {journal} {Materials Science and Engineering: A}\ }\textbf {\bibinfo {volume} {835}},\ \bibinfo {pages} {142685} (\bibinfo {year} {2022}{\natexlab{b}})}\BibitemShut {NoStop}%
\bibitem [{\citenamefont {Hadorn}\ \emph {et~al.}(2012)\citenamefont {Hadorn}, \citenamefont {Hantzsche}, \citenamefont {Yi}, \citenamefont {Bohlen}, \citenamefont {Letzig}, \citenamefont {Wollmershauser},\ and\ \citenamefont {Agnew}}]{hadorn2012role}%
  \BibitemOpen
  \bibfield  {author} {\bibinfo {author} {\bibfnamefont {J.~P.}\ \bibnamefont {Hadorn}}, \bibinfo {author} {\bibfnamefont {K.}~\bibnamefont {Hantzsche}}, \bibinfo {author} {\bibfnamefont {S.}~\bibnamefont {Yi}}, \bibinfo {author} {\bibfnamefont {J.}~\bibnamefont {Bohlen}}, \bibinfo {author} {\bibfnamefont {D.}~\bibnamefont {Letzig}}, \bibinfo {author} {\bibfnamefont {J.~A.}\ \bibnamefont {Wollmershauser}},\ and\ \bibinfo {author} {\bibfnamefont {S.~R.}\ \bibnamefont {Agnew}},\ }\bibfield  {title} {\bibinfo {title} {Role of solute in the texture modification during hot deformation of {M}g-rare earth alloys},\ }\href@noop {} {\bibfield  {journal} {\bibinfo  {journal} {Metallurgical and Materials Transactions A}\ }\textbf {\bibinfo {volume} {43}},\ \bibinfo {pages} {1347} (\bibinfo {year} {2012})}\BibitemShut {NoStop}%
\bibitem [{\citenamefont {Robson}\ \emph {et~al.}(2016)\citenamefont {Robson}, \citenamefont {Haigh}, \citenamefont {Davis},\ and\ \citenamefont {Griffiths}}]{robson2016grain}%
  \BibitemOpen
  \bibfield  {author} {\bibinfo {author} {\bibfnamefont {J.~D.}\ \bibnamefont {Robson}}, \bibinfo {author} {\bibfnamefont {S.~J.}\ \bibnamefont {Haigh}}, \bibinfo {author} {\bibfnamefont {B.}~\bibnamefont {Davis}},\ and\ \bibinfo {author} {\bibfnamefont {D.}~\bibnamefont {Griffiths}},\ }\bibfield  {title} {\bibinfo {title} {Grain boundary segregation of rare-earth elements in magnesium alloys},\ }\href@noop {} {\bibfield  {journal} {\bibinfo  {journal} {Metallurgical and Materials Transactions A}\ }\textbf {\bibinfo {volume} {47}},\ \bibinfo {pages} {522} (\bibinfo {year} {2016})}\BibitemShut {NoStop}%
\bibitem [{\citenamefont {Langelier}\ \emph {et~al.}(2017)\citenamefont {Langelier}, \citenamefont {Sha}, \citenamefont {Korinek}, \citenamefont {Donnadieu}, \citenamefont {Ringer},\ and\ \citenamefont {Esmaeili}}]{langelier2017effects}%
  \BibitemOpen
  \bibfield  {author} {\bibinfo {author} {\bibfnamefont {B.}~\bibnamefont {Langelier}}, \bibinfo {author} {\bibfnamefont {G.}~\bibnamefont {Sha}}, \bibinfo {author} {\bibfnamefont {A.}~\bibnamefont {Korinek}}, \bibinfo {author} {\bibfnamefont {P.}~\bibnamefont {Donnadieu}}, \bibinfo {author} {\bibfnamefont {S.}~\bibnamefont {Ringer}},\ and\ \bibinfo {author} {\bibfnamefont {S.}~\bibnamefont {Esmaeili}},\ }\bibfield  {title} {\bibinfo {title} {The effects of microalloying on the precipitate microstructure at grain boundary regions in an {M}g-{Z}n-based alloy},\ }\href@noop {} {\bibfield  {journal} {\bibinfo  {journal} {Materials \& Design}\ }\textbf {\bibinfo {volume} {119}},\ \bibinfo {pages} {290} (\bibinfo {year} {2017})}\BibitemShut {NoStop}%
\bibitem [{\citenamefont {Qian}\ \emph {et~al.}(2022)\citenamefont {Qian}, \citenamefont {Dong}, \citenamefont {Jiang}, \citenamefont {Lei}, \citenamefont {Yang}, \citenamefont {He}, \citenamefont {Liu}, \citenamefont {Wang}, \citenamefont {Yuan}, \citenamefont {Yang} \emph {et~al.}}]{qian2022influence}%
  \BibitemOpen
  \bibfield  {author} {\bibinfo {author} {\bibfnamefont {X.}~\bibnamefont {Qian}}, \bibinfo {author} {\bibfnamefont {Z.}~\bibnamefont {Dong}}, \bibinfo {author} {\bibfnamefont {B.}~\bibnamefont {Jiang}}, \bibinfo {author} {\bibfnamefont {B.}~\bibnamefont {Lei}}, \bibinfo {author} {\bibfnamefont {H.}~\bibnamefont {Yang}}, \bibinfo {author} {\bibfnamefont {C.}~\bibnamefont {He}}, \bibinfo {author} {\bibfnamefont {L.}~\bibnamefont {Liu}}, \bibinfo {author} {\bibfnamefont {C.}~\bibnamefont {Wang}}, \bibinfo {author} {\bibfnamefont {M.}~\bibnamefont {Yuan}}, \bibinfo {author} {\bibfnamefont {H.}~\bibnamefont {Yang}}, \emph {et~al.},\ }\bibfield  {title} {\bibinfo {title} {Influence of alloying element segregation at grain boundary on the microstructure and mechanical properties of {M}g-{Z}n alloy},\ }\href@noop {} {\bibfield  {journal} {\bibinfo  {journal} {Materials \& Design}\ }\textbf {\bibinfo {volume} {224}},\ \bibinfo {pages} {111322} (\bibinfo {year} {2022})}\BibitemShut {NoStop}%
\bibitem [{\citenamefont {Zhang}\ \emph {et~al.}(2022)\citenamefont {Zhang}, \citenamefont {Zhang}, \citenamefont {Xie}, \citenamefont {Liu}, \citenamefont {He}, \citenamefont {Wang}, \citenamefont {Fang}, \citenamefont {Fu}, \citenamefont {Jiao},\ and\ \citenamefont {Wu}}]{zhang2022significantly}%
  \BibitemOpen
  \bibfield  {author} {\bibinfo {author} {\bibfnamefont {Z.}~\bibnamefont {Zhang}}, \bibinfo {author} {\bibfnamefont {J.}~\bibnamefont {Zhang}}, \bibinfo {author} {\bibfnamefont {J.}~\bibnamefont {Xie}}, \bibinfo {author} {\bibfnamefont {S.}~\bibnamefont {Liu}}, \bibinfo {author} {\bibfnamefont {Y.}~\bibnamefont {He}}, \bibinfo {author} {\bibfnamefont {R.}~\bibnamefont {Wang}}, \bibinfo {author} {\bibfnamefont {D.}~\bibnamefont {Fang}}, \bibinfo {author} {\bibfnamefont {W.}~\bibnamefont {Fu}}, \bibinfo {author} {\bibfnamefont {Y.}~\bibnamefont {Jiao}},\ and\ \bibinfo {author} {\bibfnamefont {R.}~\bibnamefont {Wu}},\ }\bibfield  {title} {\bibinfo {title} {Significantly enhanced grain boundary {Z}n and {C}a co-segregation of dilute {M}g alloy via trace {S}m addition},\ }\href@noop {} {\bibfield  {journal} {\bibinfo  {journal} {Materials Science and Engineering: A}\ }\textbf {\bibinfo {volume} {831}},\ \bibinfo {pages} {142259} (\bibinfo {year} {2022})}\BibitemShut {NoStop}%
\bibitem [{\citenamefont {Pei}\ \emph {et~al.}(2023)\citenamefont {Pei}, \citenamefont {Xie}, \citenamefont {Yi}, \citenamefont {Korte-Kerzel}, \citenamefont {Gu{\'e}nol{\'e}},\ and\ \citenamefont {Al-Samman}}]{pei2023atomistic}%
  \BibitemOpen
  \bibfield  {author} {\bibinfo {author} {\bibfnamefont {R.}~\bibnamefont {Pei}}, \bibinfo {author} {\bibfnamefont {Z.}~\bibnamefont {Xie}}, \bibinfo {author} {\bibfnamefont {S.}~\bibnamefont {Yi}}, \bibinfo {author} {\bibfnamefont {S.}~\bibnamefont {Korte-Kerzel}}, \bibinfo {author} {\bibfnamefont {J.}~\bibnamefont {Gu{\'e}nol{\'e}}},\ and\ \bibinfo {author} {\bibfnamefont {T.}~\bibnamefont {Al-Samman}},\ }\bibfield  {title} {\bibinfo {title} {Atomistic insights into the inhomogeneous nature of solute segregation to grain boundaries in magnesium},\ }\href@noop {} {\bibfield  {journal} {\bibinfo  {journal} {Scripta Materialia}\ }\textbf {\bibinfo {volume} {230}},\ \bibinfo {pages} {115432} (\bibinfo {year} {2023})}\BibitemShut {NoStop}%
\bibitem [{\citenamefont {Mouhib}\ \emph {et~al.}(2024)\citenamefont {Mouhib}, \citenamefont {Xie}, \citenamefont {Atila}, \citenamefont {Gu{\'e}nol{\'e}}, \citenamefont {Korte-Kerzel},\ and\ \citenamefont {Al-Samman}}]{mouhib2024exploring}%
  \BibitemOpen
  \bibfield  {author} {\bibinfo {author} {\bibfnamefont {F.}~\bibnamefont {Mouhib}}, \bibinfo {author} {\bibfnamefont {Z.}~\bibnamefont {Xie}}, \bibinfo {author} {\bibfnamefont {A.}~\bibnamefont {Atila}}, \bibinfo {author} {\bibfnamefont {J.}~\bibnamefont {Gu{\'e}nol{\'e}}}, \bibinfo {author} {\bibfnamefont {S.}~\bibnamefont {Korte-Kerzel}},\ and\ \bibinfo {author} {\bibfnamefont {T.}~\bibnamefont {Al-Samman}},\ }\bibfield  {title} {\bibinfo {title} {Exploring solute behavior and texture selection in magnesium alloys at the atomistic level},\ }\href@noop {} {\bibfield  {journal} {\bibinfo  {journal} {Acta Materialia}\ }\textbf {\bibinfo {volume} {266}},\ \bibinfo {pages} {119677} (\bibinfo {year} {2024})}\BibitemShut {NoStop}%
\bibitem [{\citenamefont {Huber}\ \emph {et~al.}(2014)\citenamefont {Huber}, \citenamefont {Rottler},\ and\ \citenamefont {Militzer}}]{huber2014atomistic}%
  \BibitemOpen
  \bibfield  {author} {\bibinfo {author} {\bibfnamefont {L.}~\bibnamefont {Huber}}, \bibinfo {author} {\bibfnamefont {J.}~\bibnamefont {Rottler}},\ and\ \bibinfo {author} {\bibfnamefont {M.}~\bibnamefont {Militzer}},\ }\bibfield  {title} {\bibinfo {title} {Atomistic simulations of the interaction of alloying elements with grain boundaries in {M}g},\ }\href@noop {} {\bibfield  {journal} {\bibinfo  {journal} {Acta materialia}\ }\textbf {\bibinfo {volume} {80}},\ \bibinfo {pages} {194} (\bibinfo {year} {2014})}\BibitemShut {NoStop}%
\bibitem [{\citenamefont {Wang}\ \emph {et~al.}(2024)\citenamefont {Wang}, \citenamefont {Gu{\'e}nol{\'e}}, \citenamefont {Korte-Kerzel}, \citenamefont {Al-Samman},\ and\ \citenamefont {Xie}}]{wang2024defects}%
  \BibitemOpen
  \bibfield  {author} {\bibinfo {author} {\bibfnamefont {H.}~\bibnamefont {Wang}}, \bibinfo {author} {\bibfnamefont {J.}~\bibnamefont {Gu{\'e}nol{\'e}}}, \bibinfo {author} {\bibfnamefont {S.}~\bibnamefont {Korte-Kerzel}}, \bibinfo {author} {\bibfnamefont {T.}~\bibnamefont {Al-Samman}},\ and\ \bibinfo {author} {\bibfnamefont {Z.}~\bibnamefont {Xie}},\ }\bibfield  {title} {\bibinfo {title} {Defects in magnesium and its alloys by atomistic simulation: Assessment of semi-empirical potentials},\ }\href@noop {} {\bibfield  {journal} {\bibinfo  {journal} {Computational Materials Science}\ }\textbf {\bibinfo {volume} {240}},\ \bibinfo {pages} {113025} (\bibinfo {year} {2024})}\BibitemShut {NoStop}%
\bibitem [{\citenamefont {Pei}\ \emph {et~al.}(2019)\citenamefont {Pei}, \citenamefont {Li}, \citenamefont {Nie},\ and\ \citenamefont {Morris}}]{pei2019first}%
  \BibitemOpen
  \bibfield  {author} {\bibinfo {author} {\bibfnamefont {Z.}~\bibnamefont {Pei}}, \bibinfo {author} {\bibfnamefont {R.}~\bibnamefont {Li}}, \bibinfo {author} {\bibfnamefont {J.-F.}\ \bibnamefont {Nie}},\ and\ \bibinfo {author} {\bibfnamefont {J.~R.}\ \bibnamefont {Morris}},\ }\bibfield  {title} {\bibinfo {title} {First-principles study of the solute segregation in twin boundaries in {M}g and possible descriptors for mechanical properties},\ }\href@noop {} {\bibfield  {journal} {\bibinfo  {journal} {Materials \& Design}\ }\textbf {\bibinfo {volume} {165}},\ \bibinfo {pages} {107574} (\bibinfo {year} {2019})}\BibitemShut {NoStop}%
\bibitem [{\citenamefont {Wagih}\ \emph {et~al.}(2020)\citenamefont {Wagih}, \citenamefont {Larsen},\ and\ \citenamefont {Schuh}}]{wagih2020learning}%
  \BibitemOpen
  \bibfield  {author} {\bibinfo {author} {\bibfnamefont {M.}~\bibnamefont {Wagih}}, \bibinfo {author} {\bibfnamefont {P.~M.}\ \bibnamefont {Larsen}},\ and\ \bibinfo {author} {\bibfnamefont {C.~A.}\ \bibnamefont {Schuh}},\ }\bibfield  {title} {\bibinfo {title} {Learning grain boundary segregation energy spectra in polycrystals},\ }\href@noop {} {\bibfield  {journal} {\bibinfo  {journal} {Nature communications}\ }\textbf {\bibinfo {volume} {11}},\ \bibinfo {pages} {6376} (\bibinfo {year} {2020})}\BibitemShut {NoStop}%
\bibitem [{\citenamefont {Messina}\ \emph {et~al.}(2021)\citenamefont {Messina}, \citenamefont {Luo}, \citenamefont {Xu}, \citenamefont {Lu}, \citenamefont {Deng}, \citenamefont {Tschopp},\ and\ \citenamefont {Gao}}]{messina2021machine}%
  \BibitemOpen
  \bibfield  {author} {\bibinfo {author} {\bibfnamefont {J.}~\bibnamefont {Messina}}, \bibinfo {author} {\bibfnamefont {R.}~\bibnamefont {Luo}}, \bibinfo {author} {\bibfnamefont {K.}~\bibnamefont {Xu}}, \bibinfo {author} {\bibfnamefont {G.}~\bibnamefont {Lu}}, \bibinfo {author} {\bibfnamefont {H.}~\bibnamefont {Deng}}, \bibinfo {author} {\bibfnamefont {M.~A.}\ \bibnamefont {Tschopp}},\ and\ \bibinfo {author} {\bibfnamefont {F.}~\bibnamefont {Gao}},\ }\bibfield  {title} {\bibinfo {title} {Machine learning to predict aluminum segregation to magnesium grain boundaries},\ }\href@noop {} {\bibfield  {journal} {\bibinfo  {journal} {Scripta Materialia}\ }\textbf {\bibinfo {volume} {204}},\ \bibinfo {pages} {114150} (\bibinfo {year} {2021})}\BibitemShut {NoStop}%
\bibitem [{\citenamefont {Menon}\ \emph {et~al.}(2024)\citenamefont {Menon}, \citenamefont {Das}, \citenamefont {Gavini},\ and\ \citenamefont {Qi}}]{menon2024atomistic}%
  \BibitemOpen
  \bibfield  {author} {\bibinfo {author} {\bibfnamefont {V.}~\bibnamefont {Menon}}, \bibinfo {author} {\bibfnamefont {S.}~\bibnamefont {Das}}, \bibinfo {author} {\bibfnamefont {V.}~\bibnamefont {Gavini}},\ and\ \bibinfo {author} {\bibfnamefont {L.}~\bibnamefont {Qi}},\ }\bibfield  {title} {\bibinfo {title} {Atomistic simulations and machine learning of solute grain boundary segregation in mg alloys at finite temperatures},\ }\href@noop {} {\bibfield  {journal} {\bibinfo  {journal} {Acta Materialia}\ }\textbf {\bibinfo {volume} {264}},\ \bibinfo {pages} {119515} (\bibinfo {year} {2024})}\BibitemShut {NoStop}%
\bibitem [{\citenamefont {McLean}(1957)}]{mclean1957grain}%
  \BibitemOpen
  \bibfield  {author} {\bibinfo {author} {\bibfnamefont {D.}~\bibnamefont {McLean}},\ }\bibfield  {title} {\bibinfo {title} {Grain boundaries in metals},\ }\href@noop {} {\  (\bibinfo {year} {1957})}\BibitemShut {NoStop}%
\bibitem [{\citenamefont {White}\ and\ \citenamefont {Coghlan}(1977)}]{white1977spectrum}%
  \BibitemOpen
  \bibfield  {author} {\bibinfo {author} {\bibfnamefont {C.}~\bibnamefont {White}}\ and\ \bibinfo {author} {\bibfnamefont {W.}~\bibnamefont {Coghlan}},\ }\bibfield  {title} {\bibinfo {title} {Spectrum of binding energies approach to grain boundary segregation},\ }\href@noop {} {\bibfield  {journal} {\bibinfo  {journal} {Metall. Trans., A;(United States)}\ }\textbf {\bibinfo {volume} {8}} (\bibinfo {year} {1977})}\BibitemShut {NoStop}%
\bibitem [{\citenamefont {Wagih}\ and\ \citenamefont {Schuh}(2023)}]{wagih2023can}%
  \BibitemOpen
  \bibfield  {author} {\bibinfo {author} {\bibfnamefont {M.}~\bibnamefont {Wagih}}\ and\ \bibinfo {author} {\bibfnamefont {C.~A.}\ \bibnamefont {Schuh}},\ }\bibfield  {title} {\bibinfo {title} {Can symmetric tilt grain boundaries represent polycrystals?},\ }\href@noop {} {\bibfield  {journal} {\bibinfo  {journal} {Scripta Materialia}\ ,\ \bibinfo {pages} {115716}} (\bibinfo {year} {2023})}\BibitemShut {NoStop}%
\bibitem [{\citenamefont {Zhou}\ \emph {et~al.}(2004)\citenamefont {Zhou}, \citenamefont {Johnson},\ and\ \citenamefont {Wadley}}]{zhou2004misfit}%
  \BibitemOpen
  \bibfield  {author} {\bibinfo {author} {\bibfnamefont {X.}~\bibnamefont {Zhou}}, \bibinfo {author} {\bibfnamefont {R.}~\bibnamefont {Johnson}},\ and\ \bibinfo {author} {\bibfnamefont {H.}~\bibnamefont {Wadley}},\ }\bibfield  {title} {\bibinfo {title} {Misfit-energy-increasing dislocations in vapor-deposited {C}o{F}e/{N}i{F}e multilayers},\ }\href@noop {} {\bibfield  {journal} {\bibinfo  {journal} {Physical Review B}\ }\textbf {\bibinfo {volume} {69}},\ \bibinfo {pages} {144113} (\bibinfo {year} {2004})}\BibitemShut {NoStop}%
\bibitem [{\citenamefont {Thompson}\ \emph {et~al.}(2022)\citenamefont {Thompson}, \citenamefont {Aktulga}, \citenamefont {Berger}, \citenamefont {Bolintineanu}, \citenamefont {Brown}, \citenamefont {Crozier}, \citenamefont {In't~Veld}, \citenamefont {Kohlmeyer}, \citenamefont {Moore}, \citenamefont {Nguyen} \emph {et~al.}}]{thompson2022lammps}%
  \BibitemOpen
  \bibfield  {author} {\bibinfo {author} {\bibfnamefont {A.~P.}\ \bibnamefont {Thompson}}, \bibinfo {author} {\bibfnamefont {H.~M.}\ \bibnamefont {Aktulga}}, \bibinfo {author} {\bibfnamefont {R.}~\bibnamefont {Berger}}, \bibinfo {author} {\bibfnamefont {D.~S.}\ \bibnamefont {Bolintineanu}}, \bibinfo {author} {\bibfnamefont {W.~M.}\ \bibnamefont {Brown}}, \bibinfo {author} {\bibfnamefont {P.~S.}\ \bibnamefont {Crozier}}, \bibinfo {author} {\bibfnamefont {P.~J.}\ \bibnamefont {In't~Veld}}, \bibinfo {author} {\bibfnamefont {A.}~\bibnamefont {Kohlmeyer}}, \bibinfo {author} {\bibfnamefont {S.~G.}\ \bibnamefont {Moore}}, \bibinfo {author} {\bibfnamefont {T.~D.}\ \bibnamefont {Nguyen}}, \emph {et~al.},\ }\bibfield  {title} {\bibinfo {title} {{LAMMPS}-a flexible simulation tool for particle-based materials modeling at the atomic, meso, and continuum scales},\ }\href@noop {} {\bibfield  {journal} {\bibinfo  {journal} {Computer Physics Communications}\ }\textbf {\bibinfo {volume} {271}},\ \bibinfo {pages} {108171}
  (\bibinfo {year} {2022})}\BibitemShut {NoStop}%
\bibitem [{\citenamefont {Kim}\ and\ \citenamefont {Lee}(2017)}]{kim2017modified}%
  \BibitemOpen
  \bibfield  {author} {\bibinfo {author} {\bibfnamefont {K.-H.}\ \bibnamefont {Kim}}\ and\ \bibinfo {author} {\bibfnamefont {B.-J.}\ \bibnamefont {Lee}},\ }\bibfield  {title} {\bibinfo {title} {Modified embedded-atom method interatomic potentials for {M}g-{N}d and {M}g-{P}b binary systems},\ }\href@noop {} {\bibfield  {journal} {\bibinfo  {journal} {Calphad}\ }\textbf {\bibinfo {volume} {57}},\ \bibinfo {pages} {55} (\bibinfo {year} {2017})}\BibitemShut {NoStop}%
\bibitem [{\citenamefont {Kim}\ \emph {et~al.}(2015)\citenamefont {Kim}, \citenamefont {Jeon},\ and\ \citenamefont {Lee}}]{kim2015modified}%
  \BibitemOpen
  \bibfield  {author} {\bibinfo {author} {\bibfnamefont {K.-H.}\ \bibnamefont {Kim}}, \bibinfo {author} {\bibfnamefont {J.~B.}\ \bibnamefont {Jeon}},\ and\ \bibinfo {author} {\bibfnamefont {B.-J.}\ \bibnamefont {Lee}},\ }\bibfield  {title} {\bibinfo {title} {Modified embedded-atom method interatomic potentials for {M}g--{X} ({X}= {Y}, {S}n, {C}a) binary systems},\ }\href@noop {} {\bibfield  {journal} {\bibinfo  {journal} {Calphad}\ }\textbf {\bibinfo {volume} {48}},\ \bibinfo {pages} {27} (\bibinfo {year} {2015})}\BibitemShut {NoStop}%
\bibitem [{\citenamefont {Kim}\ \emph {et~al.}(2012)\citenamefont {Kim}, \citenamefont {Jung},\ and\ \citenamefont {Lee}}]{kim2012atomistic}%
  \BibitemOpen
  \bibfield  {author} {\bibinfo {author} {\bibfnamefont {Y.-M.}\ \bibnamefont {Kim}}, \bibinfo {author} {\bibfnamefont {I.-H.}\ \bibnamefont {Jung}},\ and\ \bibinfo {author} {\bibfnamefont {B.-J.}\ \bibnamefont {Lee}},\ }\bibfield  {title} {\bibinfo {title} {Atomistic modeling of pure {L}i and {M}g--{L}i system},\ }\href@noop {} {\bibfield  {journal} {\bibinfo  {journal} {Modelling and Simulation in Materials Science and Engineering}\ }\textbf {\bibinfo {volume} {20}},\ \bibinfo {pages} {035005} (\bibinfo {year} {2012})}\BibitemShut {NoStop}%
\bibitem [{\citenamefont {Kim}\ \emph {et~al.}(2009)\citenamefont {Kim}, \citenamefont {Kim},\ and\ \citenamefont {Lee}}]{kim2009atomistic}%
  \BibitemOpen
  \bibfield  {author} {\bibinfo {author} {\bibfnamefont {Y.-M.}\ \bibnamefont {Kim}}, \bibinfo {author} {\bibfnamefont {N.~J.}\ \bibnamefont {Kim}},\ and\ \bibinfo {author} {\bibfnamefont {B.-J.}\ \bibnamefont {Lee}},\ }\bibfield  {title} {\bibinfo {title} {Atomistic modeling of pure {M}g and {M}g--{A}l systems},\ }\href@noop {} {\bibfield  {journal} {\bibinfo  {journal} {Calphad}\ }\textbf {\bibinfo {volume} {33}},\ \bibinfo {pages} {650} (\bibinfo {year} {2009})}\BibitemShut {NoStop}%
\bibitem [{\citenamefont {Jang}\ \emph {et~al.}(2018)\citenamefont {Jang}, \citenamefont {Kim},\ and\ \citenamefont {Lee}}]{jang2018modified}%
  \BibitemOpen
  \bibfield  {author} {\bibinfo {author} {\bibfnamefont {H.-S.}\ \bibnamefont {Jang}}, \bibinfo {author} {\bibfnamefont {K.-M.}\ \bibnamefont {Kim}},\ and\ \bibinfo {author} {\bibfnamefont {B.-J.}\ \bibnamefont {Lee}},\ }\bibfield  {title} {\bibinfo {title} {Modified embedded-atom method interatomic potentials for pure {Z}n and {M}g-{Z}n binary system},\ }\href@noop {} {\bibfield  {journal} {\bibinfo  {journal} {Calphad}\ }\textbf {\bibinfo {volume} {60}},\ \bibinfo {pages} {200} (\bibinfo {year} {2018})}\BibitemShut {NoStop}%
\bibitem [{\citenamefont {Hirel}(2015)}]{hirel2015atomsk}%
  \BibitemOpen
  \bibfield  {author} {\bibinfo {author} {\bibfnamefont {P.}~\bibnamefont {Hirel}},\ }\bibfield  {title} {\bibinfo {title} {Atomsk: A tool for manipulating and converting atomic data files},\ }\href@noop {} {\bibfield  {journal} {\bibinfo  {journal} {Computer Physics Communications}\ }\textbf {\bibinfo {volume} {197}},\ \bibinfo {pages} {212} (\bibinfo {year} {2015})}\BibitemShut {NoStop}%
\bibitem [{\citenamefont {Stukowski}(2012)}]{stukowski2012structure}%
  \BibitemOpen
  \bibfield  {author} {\bibinfo {author} {\bibfnamefont {A.}~\bibnamefont {Stukowski}},\ }\bibfield  {title} {\bibinfo {title} {Structure identification methods for atomistic simulations of crystalline materials},\ }\href@noop {} {\bibfield  {journal} {\bibinfo  {journal} {Modelling and Simulation in Materials Science and Engineering}\ }\textbf {\bibinfo {volume} {20}},\ \bibinfo {pages} {045021} (\bibinfo {year} {2012})}\BibitemShut {NoStop}%
\bibitem [{\citenamefont {Hoover}(1985)}]{hoover1985canonical}%
  \BibitemOpen
  \bibfield  {author} {\bibinfo {author} {\bibfnamefont {W.~G.}\ \bibnamefont {Hoover}},\ }\bibfield  {title} {\bibinfo {title} {Canonical dynamics: Equilibrium phase-space distributions},\ }\href@noop {} {\bibfield  {journal} {\bibinfo  {journal} {Physical review A}\ }\textbf {\bibinfo {volume} {31}},\ \bibinfo {pages} {1695} (\bibinfo {year} {1985})}\BibitemShut {NoStop}%
\bibitem [{\citenamefont {Bitzek}\ \emph {et~al.}(2006)\citenamefont {Bitzek}, \citenamefont {Koskinen}, \citenamefont {G{\"a}hler}, \citenamefont {Moseler},\ and\ \citenamefont {Gumbsch}}]{bitzek2006structural}%
  \BibitemOpen
  \bibfield  {author} {\bibinfo {author} {\bibfnamefont {E.}~\bibnamefont {Bitzek}}, \bibinfo {author} {\bibfnamefont {P.}~\bibnamefont {Koskinen}}, \bibinfo {author} {\bibfnamefont {F.}~\bibnamefont {G{\"a}hler}}, \bibinfo {author} {\bibfnamefont {M.}~\bibnamefont {Moseler}},\ and\ \bibinfo {author} {\bibfnamefont {P.}~\bibnamefont {Gumbsch}},\ }\bibfield  {title} {\bibinfo {title} {Structural relaxation made simple},\ }\href@noop {} {\bibfield  {journal} {\bibinfo  {journal} {Physical review letters}\ }\textbf {\bibinfo {volume} {97}},\ \bibinfo {pages} {170201} (\bibinfo {year} {2006})}\BibitemShut {NoStop}%
\bibitem [{\citenamefont {Gu{\'e}nol{\'e}}\ \emph {et~al.}(2020)\citenamefont {Gu{\'e}nol{\'e}}, \citenamefont {N{\"o}hring}, \citenamefont {Vaid}, \citenamefont {Houll{\'e}}, \citenamefont {Xie}, \citenamefont {Prakash},\ and\ \citenamefont {Bitzek}}]{guenole2020assessment}%
  \BibitemOpen
  \bibfield  {author} {\bibinfo {author} {\bibfnamefont {J.}~\bibnamefont {Gu{\'e}nol{\'e}}}, \bibinfo {author} {\bibfnamefont {W.~G.}\ \bibnamefont {N{\"o}hring}}, \bibinfo {author} {\bibfnamefont {A.}~\bibnamefont {Vaid}}, \bibinfo {author} {\bibfnamefont {F.}~\bibnamefont {Houll{\'e}}}, \bibinfo {author} {\bibfnamefont {Z.}~\bibnamefont {Xie}}, \bibinfo {author} {\bibfnamefont {A.}~\bibnamefont {Prakash}},\ and\ \bibinfo {author} {\bibfnamefont {E.}~\bibnamefont {Bitzek}},\ }\bibfield  {title} {\bibinfo {title} {Assessment and optimization of the fast inertial relaxation engine (fire) for energy minimization in atomistic simulations and its implementation in lammps},\ }\href@noop {} {\bibfield  {journal} {\bibinfo  {journal} {Computational Materials Science}\ }\textbf {\bibinfo {volume} {175}},\ \bibinfo {pages} {109584} (\bibinfo {year} {2020})}\BibitemShut {NoStop}%
\bibitem [{\citenamefont {Ding}\ \emph {et~al.}(2016)\citenamefont {Ding}, \citenamefont {Cheng}, \citenamefont {Sheng}, \citenamefont {Asta}, \citenamefont {Ritchie},\ and\ \citenamefont {Ma}}]{ding2016universal}%
  \BibitemOpen
  \bibfield  {author} {\bibinfo {author} {\bibfnamefont {J.}~\bibnamefont {Ding}}, \bibinfo {author} {\bibfnamefont {Y.-Q.}\ \bibnamefont {Cheng}}, \bibinfo {author} {\bibfnamefont {H.}~\bibnamefont {Sheng}}, \bibinfo {author} {\bibfnamefont {M.}~\bibnamefont {Asta}}, \bibinfo {author} {\bibfnamefont {R.~O.}\ \bibnamefont {Ritchie}},\ and\ \bibinfo {author} {\bibfnamefont {E.}~\bibnamefont {Ma}},\ }\bibfield  {title} {\bibinfo {title} {Universal structural parameter to quantitatively predict metallic glass properties},\ }\href@noop {} {\bibfield  {journal} {\bibinfo  {journal} {Nature communications}\ }\textbf {\bibinfo {volume} {7}},\ \bibinfo {pages} {13733} (\bibinfo {year} {2016})}\BibitemShut {NoStop}%
\bibitem [{\citenamefont {Creuze}\ \emph {et~al.}(2000)\citenamefont {Creuze}, \citenamefont {Berthier}, \citenamefont {T{\'e}tot}, \citenamefont {Legrand},\ and\ \citenamefont {Tr{\'e}glia}}]{creuze2000intergranular}%
  \BibitemOpen
  \bibfield  {author} {\bibinfo {author} {\bibfnamefont {J.}~\bibnamefont {Creuze}}, \bibinfo {author} {\bibfnamefont {F.}~\bibnamefont {Berthier}}, \bibinfo {author} {\bibfnamefont {R.}~\bibnamefont {T{\'e}tot}}, \bibinfo {author} {\bibfnamefont {B.}~\bibnamefont {Legrand}},\ and\ \bibinfo {author} {\bibfnamefont {G.}~\bibnamefont {Tr{\'e}glia}},\ }\bibfield  {title} {\bibinfo {title} {Intergranular segregation and vibrational effects: A local analysis},\ }\href@noop {} {\bibfield  {journal} {\bibinfo  {journal} {Physical Review B}\ }\textbf {\bibinfo {volume} {61}},\ \bibinfo {pages} {14470} (\bibinfo {year} {2000})}\BibitemShut {NoStop}%
\bibitem [{\citenamefont {Tuchinda}\ and\ \citenamefont {Schuh}(2023)}]{tuchinda2023vibrational}%
  \BibitemOpen
  \bibfield  {author} {\bibinfo {author} {\bibfnamefont {N.}~\bibnamefont {Tuchinda}}\ and\ \bibinfo {author} {\bibfnamefont {C.~A.}\ \bibnamefont {Schuh}},\ }\bibfield  {title} {\bibinfo {title} {The vibrational entropy spectra of grain boundary segregation in polycrystals},\ }\href@noop {} {\bibfield  {journal} {\bibinfo  {journal} {Acta Materialia}\ }\textbf {\bibinfo {volume} {245}},\ \bibinfo {pages} {118630} (\bibinfo {year} {2023})}\BibitemShut {NoStop}%
\bibitem [{\citenamefont {Bart{\'o}k}\ \emph {et~al.}(2010)\citenamefont {Bart{\'o}k}, \citenamefont {Payne}, \citenamefont {Kondor},\ and\ \citenamefont {Cs{\'a}nyi}}]{bartok2010gaussian}%
  \BibitemOpen
  \bibfield  {author} {\bibinfo {author} {\bibfnamefont {A.~P.}\ \bibnamefont {Bart{\'o}k}}, \bibinfo {author} {\bibfnamefont {M.~C.}\ \bibnamefont {Payne}}, \bibinfo {author} {\bibfnamefont {R.}~\bibnamefont {Kondor}},\ and\ \bibinfo {author} {\bibfnamefont {G.}~\bibnamefont {Cs{\'a}nyi}},\ }\bibfield  {title} {\bibinfo {title} {Gaussian approximation potentials: The accuracy of quantum mechanics, without the electrons},\ }\href@noop {} {\bibfield  {journal} {\bibinfo  {journal} {Physical review letters}\ }\textbf {\bibinfo {volume} {104}},\ \bibinfo {pages} {136403} (\bibinfo {year} {2010})}\BibitemShut {NoStop}%
\bibitem [{\citenamefont {Bart{\'o}k}\ \emph {et~al.}(2013)\citenamefont {Bart{\'o}k}, \citenamefont {Kondor},\ and\ \citenamefont {Cs{\'a}nyi}}]{bartok2013representing}%
  \BibitemOpen
  \bibfield  {author} {\bibinfo {author} {\bibfnamefont {A.~P.}\ \bibnamefont {Bart{\'o}k}}, \bibinfo {author} {\bibfnamefont {R.}~\bibnamefont {Kondor}},\ and\ \bibinfo {author} {\bibfnamefont {G.}~\bibnamefont {Cs{\'a}nyi}},\ }\bibfield  {title} {\bibinfo {title} {On representing chemical environments},\ }\href@noop {} {\bibfield  {journal} {\bibinfo  {journal} {Physical Review B}\ }\textbf {\bibinfo {volume} {87}},\ \bibinfo {pages} {184115} (\bibinfo {year} {2013})}\BibitemShut {NoStop}%
\bibitem [{\citenamefont {Prokhorenkova}\ \emph {et~al.}(2018)\citenamefont {Prokhorenkova}, \citenamefont {Gusev}, \citenamefont {Vorobev}, \citenamefont {Dorogush},\ and\ \citenamefont {Gulin}}]{prokhorenkova2018catboost}%
  \BibitemOpen
  \bibfield  {author} {\bibinfo {author} {\bibfnamefont {L.}~\bibnamefont {Prokhorenkova}}, \bibinfo {author} {\bibfnamefont {G.}~\bibnamefont {Gusev}}, \bibinfo {author} {\bibfnamefont {A.}~\bibnamefont {Vorobev}}, \bibinfo {author} {\bibfnamefont {A.~V.}\ \bibnamefont {Dorogush}},\ and\ \bibinfo {author} {\bibfnamefont {A.}~\bibnamefont {Gulin}},\ }\bibfield  {title} {\bibinfo {title} {Catboost: unbiased boosting with categorical features},\ }\href@noop {} {\bibfield  {journal} {\bibinfo  {journal} {Advances in neural information processing systems}\ }\textbf {\bibinfo {volume} {31}} (\bibinfo {year} {2018})}\BibitemShut {NoStop}%
\bibitem [{\citenamefont {Friedman}(2001)}]{friedman2001greedy}%
  \BibitemOpen
  \bibfield  {author} {\bibinfo {author} {\bibfnamefont {J.~H.}\ \bibnamefont {Friedman}},\ }\bibfield  {title} {\bibinfo {title} {Greedy function approximation: a gradient boosting machine},\ }\href@noop {} {\bibfield  {journal} {\bibinfo  {journal} {Annals of statistics}\ ,\ \bibinfo {pages} {1189}} (\bibinfo {year} {2001})}\BibitemShut {NoStop}%
\bibitem [{\citenamefont {Chen}\ and\ \citenamefont {Guestrin}(2016)}]{chen2016xgboost}%
  \BibitemOpen
  \bibfield  {author} {\bibinfo {author} {\bibfnamefont {T.}~\bibnamefont {Chen}}\ and\ \bibinfo {author} {\bibfnamefont {C.}~\bibnamefont {Guestrin}},\ }\bibfield  {title} {\bibinfo {title} {Xgboost: A scalable tree boosting system},\ }in\ \href@noop {} {\emph {\bibinfo {booktitle} {Proceedings of the 22nd acm sigkdd international conference on knowledge discovery and data mining}}}\ (\bibinfo {year} {2016})\ pp.\ \bibinfo {pages} {785--794}\BibitemShut {NoStop}%
\bibitem [{\citenamefont {Zou}\ and\ \citenamefont {Hastie}(2005)}]{zou2005regularization}%
  \BibitemOpen
  \bibfield  {author} {\bibinfo {author} {\bibfnamefont {H.}~\bibnamefont {Zou}}\ and\ \bibinfo {author} {\bibfnamefont {T.}~\bibnamefont {Hastie}},\ }\bibfield  {title} {\bibinfo {title} {Regularization and variable selection via the elastic net},\ }\href@noop {} {\bibfield  {journal} {\bibinfo  {journal} {Journal of the Royal Statistical Society Series B: Statistical Methodology}\ }\textbf {\bibinfo {volume} {67}},\ \bibinfo {pages} {301} (\bibinfo {year} {2005})}\BibitemShut {NoStop}%
\bibitem [{\citenamefont {Pedregosa}\ \emph {et~al.}(2011)\citenamefont {Pedregosa}, \citenamefont {Varoquaux}, \citenamefont {Gramfort}, \citenamefont {Michel}, \citenamefont {Thirion}, \citenamefont {Grisel}, \citenamefont {Blondel}, \citenamefont {Prettenhofer}, \citenamefont {Weiss}, \citenamefont {Dubourg} \emph {et~al.}}]{pedregosa2011scikit}%
  \BibitemOpen
  \bibfield  {author} {\bibinfo {author} {\bibfnamefont {F.}~\bibnamefont {Pedregosa}}, \bibinfo {author} {\bibfnamefont {G.}~\bibnamefont {Varoquaux}}, \bibinfo {author} {\bibfnamefont {A.}~\bibnamefont {Gramfort}}, \bibinfo {author} {\bibfnamefont {V.}~\bibnamefont {Michel}}, \bibinfo {author} {\bibfnamefont {B.}~\bibnamefont {Thirion}}, \bibinfo {author} {\bibfnamefont {O.}~\bibnamefont {Grisel}}, \bibinfo {author} {\bibfnamefont {M.}~\bibnamefont {Blondel}}, \bibinfo {author} {\bibfnamefont {P.}~\bibnamefont {Prettenhofer}}, \bibinfo {author} {\bibfnamefont {R.}~\bibnamefont {Weiss}}, \bibinfo {author} {\bibfnamefont {V.}~\bibnamefont {Dubourg}}, \emph {et~al.},\ }\bibfield  {title} {\bibinfo {title} {Scikit-learn: Machine learning in {P}ython},\ }\href@noop {} {\bibfield  {journal} {\bibinfo  {journal} {the Journal of machine Learning research}\ }\textbf {\bibinfo {volume} {12}},\ \bibinfo {pages} {2825} (\bibinfo {year} {2011})}\BibitemShut {NoStop}%
\bibitem [{\citenamefont {Nie}\ and\ \citenamefont {Xie}(2007)}]{nie2007ab}%
  \BibitemOpen
  \bibfield  {author} {\bibinfo {author} {\bibfnamefont {Y.}~\bibnamefont {Nie}}\ and\ \bibinfo {author} {\bibfnamefont {Y.}~\bibnamefont {Xie}},\ }\bibfield  {title} {\bibinfo {title} {Ab initio thermodynamics of the hcp metals {M}g, {T}i, and {Z}r},\ }\href@noop {} {\bibfield  {journal} {\bibinfo  {journal} {Physical Review B}\ }\textbf {\bibinfo {volume} {75}},\ \bibinfo {pages} {174117} (\bibinfo {year} {2007})}\BibitemShut {NoStop}%
\bibitem [{\citenamefont {Hultgren}\ \emph {et~al.}(1973)\citenamefont {Hultgren}, \citenamefont {Desai}, \citenamefont {Hawkins}, \citenamefont {Gleiser}, \citenamefont {Kelley},\ and\ \citenamefont {Wagman}}]{hultgren1973selected}%
  \BibitemOpen
  \bibfield  {author} {\bibinfo {author} {\bibfnamefont {R.~R.}\ \bibnamefont {Hultgren}}, \bibinfo {author} {\bibfnamefont {P.~D.}\ \bibnamefont {Desai}}, \bibinfo {author} {\bibfnamefont {D.~T.}\ \bibnamefont {Hawkins}}, \bibinfo {author} {\bibfnamefont {M.}~\bibnamefont {Gleiser}}, \bibinfo {author} {\bibfnamefont {K.~K.}\ \bibnamefont {Kelley}},\ and\ \bibinfo {author} {\bibfnamefont {D.~D.}\ \bibnamefont {Wagman}},\ }\bibfield  {title} {\bibinfo {title} {Selected values of the thermodynamic properties of the elements},\ }\href@noop {} {\  (\bibinfo {year} {1973})}\BibitemShut {NoStop}%
\bibitem [{\citenamefont {Wagih}\ and\ \citenamefont {Schuh}(2019)}]{wagih2019spectrum}%
  \BibitemOpen
  \bibfield  {author} {\bibinfo {author} {\bibfnamefont {M.}~\bibnamefont {Wagih}}\ and\ \bibinfo {author} {\bibfnamefont {C.~A.}\ \bibnamefont {Schuh}},\ }\bibfield  {title} {\bibinfo {title} {Spectrum of grain boundary segregation energies in a polycrystal},\ }\href@noop {} {\bibfield  {journal} {\bibinfo  {journal} {Acta Materialia}\ }\textbf {\bibinfo {volume} {181}},\ \bibinfo {pages} {228} (\bibinfo {year} {2019})}\BibitemShut {NoStop}%
\bibitem [{\citenamefont {Homer}\ \emph {et~al.}(2022)\citenamefont {Homer}, \citenamefont {Hart}, \citenamefont {Owens}, \citenamefont {Hensley}, \citenamefont {Spendlove},\ and\ \citenamefont {Serafin}}]{homer2022examination}%
  \BibitemOpen
  \bibfield  {author} {\bibinfo {author} {\bibfnamefont {E.~R.}\ \bibnamefont {Homer}}, \bibinfo {author} {\bibfnamefont {G.~L.}\ \bibnamefont {Hart}}, \bibinfo {author} {\bibfnamefont {C.~B.}\ \bibnamefont {Owens}}, \bibinfo {author} {\bibfnamefont {D.~M.}\ \bibnamefont {Hensley}}, \bibinfo {author} {\bibfnamefont {J.~C.}\ \bibnamefont {Spendlove}},\ and\ \bibinfo {author} {\bibfnamefont {L.~H.}\ \bibnamefont {Serafin}},\ }\bibfield  {title} {\bibinfo {title} {Examination of computed aluminum grain boundary structures and energies that span the 5d space of crystallographic character},\ }\href@noop {} {\bibfield  {journal} {\bibinfo  {journal} {Acta Materialia}\ }\textbf {\bibinfo {volume} {234}},\ \bibinfo {pages} {118006} (\bibinfo {year} {2022})}\BibitemShut {NoStop}%
\bibitem [{\citenamefont {Fujii}\ \emph {et~al.}(2020)\citenamefont {Fujii}, \citenamefont {Yokoi}, \citenamefont {Fisher}, \citenamefont {Moriwake},\ and\ \citenamefont {Yoshiya}}]{fujii2020quantitative}%
  \BibitemOpen
  \bibfield  {author} {\bibinfo {author} {\bibfnamefont {S.}~\bibnamefont {Fujii}}, \bibinfo {author} {\bibfnamefont {T.}~\bibnamefont {Yokoi}}, \bibinfo {author} {\bibfnamefont {C.~A.}\ \bibnamefont {Fisher}}, \bibinfo {author} {\bibfnamefont {H.}~\bibnamefont {Moriwake}},\ and\ \bibinfo {author} {\bibfnamefont {M.}~\bibnamefont {Yoshiya}},\ }\bibfield  {title} {\bibinfo {title} {Quantitative prediction of grain boundary thermal conductivities from local atomic environments},\ }\href@noop {} {\bibfield  {journal} {\bibinfo  {journal} {Nature communications}\ }\textbf {\bibinfo {volume} {11}},\ \bibinfo {pages} {1854} (\bibinfo {year} {2020})}\BibitemShut {NoStop}%
\bibitem [{\citenamefont {Huber}\ \emph {et~al.}(2018)\citenamefont {Huber}, \citenamefont {Hadian}, \citenamefont {Grabowski},\ and\ \citenamefont {Neugebauer}}]{huber2018machine}%
  \BibitemOpen
  \bibfield  {author} {\bibinfo {author} {\bibfnamefont {L.}~\bibnamefont {Huber}}, \bibinfo {author} {\bibfnamefont {R.}~\bibnamefont {Hadian}}, \bibinfo {author} {\bibfnamefont {B.}~\bibnamefont {Grabowski}},\ and\ \bibinfo {author} {\bibfnamefont {J.}~\bibnamefont {Neugebauer}},\ }\bibfield  {title} {\bibinfo {title} {A machine learning approach to model solute grain boundary segregation},\ }\href@noop {} {\bibfield  {journal} {\bibinfo  {journal} {npj Computational Materials}\ }\textbf {\bibinfo {volume} {4}},\ \bibinfo {pages} {64} (\bibinfo {year} {2018})}\BibitemShut {NoStop}%
\bibitem [{\citenamefont {Mahmood}\ \emph {et~al.}(2022)\citenamefont {Mahmood}, \citenamefont {Alghalayini}, \citenamefont {Martinez}, \citenamefont {Paredis},\ and\ \citenamefont {Abdeljawad}}]{mahmood2022atomistic}%
  \BibitemOpen
  \bibfield  {author} {\bibinfo {author} {\bibfnamefont {Y.}~\bibnamefont {Mahmood}}, \bibinfo {author} {\bibfnamefont {M.}~\bibnamefont {Alghalayini}}, \bibinfo {author} {\bibfnamefont {E.}~\bibnamefont {Martinez}}, \bibinfo {author} {\bibfnamefont {C.~J.}\ \bibnamefont {Paredis}},\ and\ \bibinfo {author} {\bibfnamefont {F.}~\bibnamefont {Abdeljawad}},\ }\bibfield  {title} {\bibinfo {title} {Atomistic and machine learning studies of solute segregation in metastable grain boundaries},\ }\href@noop {} {\bibfield  {journal} {\bibinfo  {journal} {Scientific Reports}\ }\textbf {\bibinfo {volume} {12}},\ \bibinfo {pages} {6673} (\bibinfo {year} {2022})}\BibitemShut {NoStop}%
\bibitem [{\citenamefont {Ma}\ and\ \citenamefont {Pan}(2023)}]{ma2023efficient}%
  \BibitemOpen
  \bibfield  {author} {\bibinfo {author} {\bibfnamefont {Z.}~\bibnamefont {Ma}}\ and\ \bibinfo {author} {\bibfnamefont {Z.}~\bibnamefont {Pan}},\ }\bibfield  {title} {\bibinfo {title} {Efficient machine learning of solute segregation energy based on physics-informed features},\ }\href@noop {} {\bibfield  {journal} {\bibinfo  {journal} {Scientific Reports}\ }\textbf {\bibinfo {volume} {13}},\ \bibinfo {pages} {11449} (\bibinfo {year} {2023})}\BibitemShut {NoStop}%
\bibitem [{\citenamefont {Borges}\ \emph {et~al.}(2024)\citenamefont {Borges}, \citenamefont {Huber}, \citenamefont {Zapolsky}, \citenamefont {Patte},\ and\ \citenamefont {Demange}}]{borges2024insights}%
  \BibitemOpen
  \bibfield  {author} {\bibinfo {author} {\bibfnamefont {Y.}~\bibnamefont {Borges}}, \bibinfo {author} {\bibfnamefont {L.}~\bibnamefont {Huber}}, \bibinfo {author} {\bibfnamefont {H.}~\bibnamefont {Zapolsky}}, \bibinfo {author} {\bibfnamefont {R.}~\bibnamefont {Patte}},\ and\ \bibinfo {author} {\bibfnamefont {G.}~\bibnamefont {Demange}},\ }\bibfield  {title} {\bibinfo {title} {Insights from symmetry: Improving machine-learned models for grain boundary segregation},\ }\href@noop {} {\bibfield  {journal} {\bibinfo  {journal} {Computational Materials Science}\ }\textbf {\bibinfo {volume} {232}},\ \bibinfo {pages} {112663} (\bibinfo {year} {2024})}\BibitemShut {NoStop}%
\bibitem [{\citenamefont {Rittner}\ and\ \citenamefont {Seidman}(1997)}]{rittner1997solute}%
  \BibitemOpen
  \bibfield  {author} {\bibinfo {author} {\bibfnamefont {J.}~\bibnamefont {Rittner}}\ and\ \bibinfo {author} {\bibfnamefont {D.}~\bibnamefont {Seidman}},\ }\bibfield  {title} {\bibinfo {title} {Solute-atom segregation to $\langle$110$\rangle$ symmetric tilt grain boundaries},\ }\href@noop {} {\bibfield  {journal} {\bibinfo  {journal} {Acta materialia}\ }\textbf {\bibinfo {volume} {45}},\ \bibinfo {pages} {3191} (\bibinfo {year} {1997})}\BibitemShut {NoStop}%
\bibitem [{\citenamefont {Lej{\v{c}}ek}\ and\ \citenamefont {Hofmann}(1995)}]{lejvcek1995thermodynamics}%
  \BibitemOpen
  \bibfield  {author} {\bibinfo {author} {\bibfnamefont {P.}~\bibnamefont {Lej{\v{c}}ek}}\ and\ \bibinfo {author} {\bibfnamefont {S.}~\bibnamefont {Hofmann}},\ }\bibfield  {title} {\bibinfo {title} {Thermodynamics and structural aspects of grain boundary segregation},\ }\href@noop {} {\bibfield  {journal} {\bibinfo  {journal} {Critical Reviews in Solid State and Material Sciences}\ }\textbf {\bibinfo {volume} {20}},\ \bibinfo {pages} {1} (\bibinfo {year} {1995})}\BibitemShut {NoStop}%
\bibitem [{\citenamefont {Chang}\ \emph {et~al.}(2006)\citenamefont {Chang}, \citenamefont {Wang}, \citenamefont {Chu},\ and\ \citenamefont {Lee}}]{chang2006mechanical}%
  \BibitemOpen
  \bibfield  {author} {\bibinfo {author} {\bibfnamefont {T.-C.}\ \bibnamefont {Chang}}, \bibinfo {author} {\bibfnamefont {J.-Y.}\ \bibnamefont {Wang}}, \bibinfo {author} {\bibfnamefont {C.-L.}\ \bibnamefont {Chu}},\ and\ \bibinfo {author} {\bibfnamefont {S.}~\bibnamefont {Lee}},\ }\bibfield  {title} {\bibinfo {title} {Mechanical properties and microstructures of various {M}g--{L}i alloys},\ }\href@noop {} {\bibfield  {journal} {\bibinfo  {journal} {Materials Letters}\ }\textbf {\bibinfo {volume} {60}},\ \bibinfo {pages} {3272} (\bibinfo {year} {2006})}\BibitemShut {NoStop}%
\bibitem [{\citenamefont {Peng}\ \emph {et~al.}(2022)\citenamefont {Peng}, \citenamefont {Liu}, \citenamefont {Wu}, \citenamefont {Ji},\ and\ \citenamefont {Ding}}]{peng2022plastic}%
  \BibitemOpen
  \bibfield  {author} {\bibinfo {author} {\bibfnamefont {X.}~\bibnamefont {Peng}}, \bibinfo {author} {\bibfnamefont {W.}~\bibnamefont {Liu}}, \bibinfo {author} {\bibfnamefont {G.}~\bibnamefont {Wu}}, \bibinfo {author} {\bibfnamefont {H.}~\bibnamefont {Ji}},\ and\ \bibinfo {author} {\bibfnamefont {W.}~\bibnamefont {Ding}},\ }\bibfield  {title} {\bibinfo {title} {Plastic deformation and heat treatment of {M}g-{L}i alloys: a review},\ }\href@noop {} {\bibfield  {journal} {\bibinfo  {journal} {Journal of Materials Science \& Technology}\ }\textbf {\bibinfo {volume} {99}},\ \bibinfo {pages} {193} (\bibinfo {year} {2022})}\BibitemShut {NoStop}%
\bibitem [{\citenamefont {Sadigh}\ \emph {et~al.}(2012)\citenamefont {Sadigh}, \citenamefont {Erhart}, \citenamefont {Stukowski}, \citenamefont {Caro}, \citenamefont {Martinez},\ and\ \citenamefont {Zepeda-Ruiz}}]{sadigh2012scalable}%
  \BibitemOpen
  \bibfield  {author} {\bibinfo {author} {\bibfnamefont {B.}~\bibnamefont {Sadigh}}, \bibinfo {author} {\bibfnamefont {P.}~\bibnamefont {Erhart}}, \bibinfo {author} {\bibfnamefont {A.}~\bibnamefont {Stukowski}}, \bibinfo {author} {\bibfnamefont {A.}~\bibnamefont {Caro}}, \bibinfo {author} {\bibfnamefont {E.}~\bibnamefont {Martinez}},\ and\ \bibinfo {author} {\bibfnamefont {L.}~\bibnamefont {Zepeda-Ruiz}},\ }\bibfield  {title} {\bibinfo {title} {Scalable parallel monte carlo algorithm for atomistic simulations of precipitation in alloys},\ }\href@noop {} {\bibfield  {journal} {\bibinfo  {journal} {Physical Review B}\ }\textbf {\bibinfo {volume} {85}},\ \bibinfo {pages} {184203} (\bibinfo {year} {2012})}\BibitemShut {NoStop}%
\bibitem [{\citenamefont {Pan}\ and\ \citenamefont {Rupert}(2016)}]{pan2016effect}%
  \BibitemOpen
  \bibfield  {author} {\bibinfo {author} {\bibfnamefont {Z.}~\bibnamefont {Pan}}\ and\ \bibinfo {author} {\bibfnamefont {T.~J.}\ \bibnamefont {Rupert}},\ }\bibfield  {title} {\bibinfo {title} {Effect of grain boundary character on segregation-induced structural transitions},\ }\href@noop {} {\bibfield  {journal} {\bibinfo  {journal} {Physical Review B}\ }\textbf {\bibinfo {volume} {93}},\ \bibinfo {pages} {134113} (\bibinfo {year} {2016})}\BibitemShut {NoStop}%
\bibitem [{\citenamefont {Ganguly}\ \emph {et~al.}(2024)\citenamefont {Ganguly}, \citenamefont {Wang}, \citenamefont {Gu{\'e}nol{\'e}}, \citenamefont {Prakash}, \citenamefont {Korte-Kerzel}, \citenamefont {Al-Samman},\ and\ \citenamefont {Xie}}]{ganguly2024grain}%
  \BibitemOpen
  \bibfield  {author} {\bibinfo {author} {\bibfnamefont {A.}~\bibnamefont {Ganguly}}, \bibinfo {author} {\bibfnamefont {H.}~\bibnamefont {Wang}}, \bibinfo {author} {\bibfnamefont {J.}~\bibnamefont {Gu{\'e}nol{\'e}}}, \bibinfo {author} {\bibfnamefont {A.}~\bibnamefont {Prakash}}, \bibinfo {author} {\bibfnamefont {S.}~\bibnamefont {Korte-Kerzel}}, \bibinfo {author} {\bibfnamefont {T.}~\bibnamefont {Al-Samman}},\ and\ \bibinfo {author} {\bibfnamefont {Z.}~\bibnamefont {Xie}},\ }\bibfield  {title} {\bibinfo {title} {Grain boundary segregation spectrum in basal-textured {M}g alloys: From solute decoration to structural transition},\ }\href@noop {} {\bibfield  {journal} {\bibinfo  {journal} {Acta Materialia}\ ,\ \bibinfo {pages} {120556}} (\bibinfo {year} {2024})}\BibitemShut {NoStop}%
\bibitem [{\citenamefont {Korte-Kerzel}\ \emph {et~al.}(2022)\citenamefont {Korte-Kerzel}, \citenamefont {Hickel}, \citenamefont {Huber}, \citenamefont {Raabe}, \citenamefont {Sandl{\"o}bes-Haut}, \citenamefont {Todorova},\ and\ \citenamefont {Neugebauer}}]{korte2022defect}%
  \BibitemOpen
  \bibfield  {author} {\bibinfo {author} {\bibfnamefont {S.}~\bibnamefont {Korte-Kerzel}}, \bibinfo {author} {\bibfnamefont {T.}~\bibnamefont {Hickel}}, \bibinfo {author} {\bibfnamefont {L.}~\bibnamefont {Huber}}, \bibinfo {author} {\bibfnamefont {D.}~\bibnamefont {Raabe}}, \bibinfo {author} {\bibfnamefont {S.}~\bibnamefont {Sandl{\"o}bes-Haut}}, \bibinfo {author} {\bibfnamefont {M.}~\bibnamefont {Todorova}},\ and\ \bibinfo {author} {\bibfnamefont {J.}~\bibnamefont {Neugebauer}},\ }\bibfield  {title} {\bibinfo {title} {Defect phases--thermodynamics and impact on material properties},\ }\href@noop {} {\bibfield  {journal} {\bibinfo  {journal} {International Materials Reviews}\ }\textbf {\bibinfo {volume} {67}},\ \bibinfo {pages} {89} (\bibinfo {year} {2022})}\BibitemShut {NoStop}%
\bibitem [{\citenamefont {Mendelev}\ \emph {et~al.}(2009)\citenamefont {Mendelev}, \citenamefont {Asta}, \citenamefont {Rahman},\ and\ \citenamefont {Hoyt}}]{mendelev2009development}%
  \BibitemOpen
  \bibfield  {author} {\bibinfo {author} {\bibfnamefont {M.}~\bibnamefont {Mendelev}}, \bibinfo {author} {\bibfnamefont {M.}~\bibnamefont {Asta}}, \bibinfo {author} {\bibfnamefont {M.}~\bibnamefont {Rahman}},\ and\ \bibinfo {author} {\bibfnamefont {J.}~\bibnamefont {Hoyt}},\ }\bibfield  {title} {\bibinfo {title} {Development of interatomic potentials appropriate for simulation of solid--liquid interface properties in {A}l--{M}g alloys},\ }\href@noop {} {\bibfield  {journal} {\bibinfo  {journal} {Philosophical Magazine}\ }\textbf {\bibinfo {volume} {89}},\ \bibinfo {pages} {3269} (\bibinfo {year} {2009})}\BibitemShut {NoStop}%
\bibitem [{\citenamefont {Wagih}\ and\ \citenamefont {Schuh}(2022)}]{wagih2022learning}%
  \BibitemOpen
  \bibfield  {author} {\bibinfo {author} {\bibfnamefont {M.}~\bibnamefont {Wagih}}\ and\ \bibinfo {author} {\bibfnamefont {C.~A.}\ \bibnamefont {Schuh}},\ }\bibfield  {title} {\bibinfo {title} {Learning grain-boundary segregation: from first principles to polycrystals},\ }\href@noop {} {\bibfield  {journal} {\bibinfo  {journal} {Physical review letters}\ }\textbf {\bibinfo {volume} {129}},\ \bibinfo {pages} {046102} (\bibinfo {year} {2022})}\BibitemShut {NoStop}%
\end{thebibliography}%

\end{document}